\documentclass{article}
\usepackage{graphicx}
\usepackage{pdfpages,epstopdf}
\DeclareGraphicsExtensions{.pdf}
\newcommand{\bc}{\begin{center}}
\newcommand{\ec}{\end{center}}
\newcommand{\bdm}{\begin{displaymath}}
\newcommand{\edm}{\end{displaymath}}
\newcommand{\be}{\begin{equation}}
\newcommand{\ee}{\end{equation}}
\newcommand{\bea}{\begin{eqnarray}}
\newcommand{\eea}{\end{eqnarray}}
\newcommand{\bfg}{\begin{figure}[p]}
\newcommand{\efg}{\end{figure}}
\newcommand{\ta}{\ensuremath{\Theta}}
\newcommand{\tao}{\ensuremath{\Theta_1}}
\newcommand{\tp}{\ensuremath{\Theta^{\prime}}}
\newcommand{\tpp}{\ensuremath{\Theta^{\prime \prime}}}

\newcommand{\ph}{\ensuremath{\frac{\pi}{2}}}
\newcommand{\phtext}{\ensuremath{\pi/2}}
\newcommand{\DP}{\ensuremath{\Delta\Phi}}
\newcommand{\vect}[1]{\ensuremath{\mbox{\bf #1}}}
\newcommand{\unitv}[1]{\ensuremath{\mbox{\textbf {\^{#1}}}}}
\newcommand{\vind}[2]{\ensuremath{\mbox{\textbf {#1}}_{#2}}}
\newcommand{\uvind}[2]{\ensuremath{\mbox{\textbf{\^{x}}}_{#2}}}
\newcommand{\eas}{equiangular spiral }
\newcommand{\abs}[1]{\mid #1 \mid}

\begin{document}

\bibliographystyle{abbrv}

\title{Closed Light Paths in Equiangular Spiral Disks}
  \author{E. Hitzer~\footnote{Dep. of Mech. Engineering, Fukui Univ., 
          Bunkyo 3-9-1, 910-8507 Fukui, Japan, hitzer@mech.fukui-u.ac.jp}}
  \date{13 September 2000}
\maketitle

\begin{abstract}

A new type of deformation for microscopic laser disks, the \textit{equiangular 
spiral deformation} is proposed. First a short review of the geometry of light 
paths in equiangular spirals in the language of real two-dimensional geometric 
calculus is given. Second, the constituting equations for 
\textit{closed paths} inside equiangular spirals are derived. 
Third, their numerical solution is performed and found to yield two 
generic types of closed light paths. \textit{Degenerate} closed paths that 
exist over large intervals of the deformation parameter, and 
\textit{nondegenerate} closed paths which only exist over relatively small 
deformation parameter intervals spanning less than 1\% of the nondegenerate 
intervals. Fourth, amongst the nondegenerate paths a 
\textit{stable asymmetric bow-tie} shaped light trajectory was found. 

\end{abstract}

\section{Introduction}

Quantum cascade laser disks deformed into \textit{flattened} 
quadrupoles~\cite{bowtie} or ovals~\cite{Laser-Flash,JN:WWW,Phot-Bill} 
have been shown to exhibit a quasi-exponential increase of the collected 
emitted power with increasing deformation. At the same time such 
micro-lasers provide a dramatic increase in directionality. 
In the favourable directions a power increase of up to three orders was 
obtained. The micro-disks are manufactured from high refractive index 
($ n_r \approx 3.3 $) materials like as InGaAs/InAlAs systems. At small 
deformations chaotic \textit{whispering gallery} modes dominate, being 
replaced by stable symmetric \textit{bow-tie} resonator modes at higher 
deformations~\cite{bowtie}. 

Theoretical considerations show that deformation from circularity causes 
the wave equation to be inseparable and the solution can no longer be indexed 
by quantum numbers. A new theoretical approach providing better physical 
understanding is the study of the short wavelength limit of the problem, 
i.e., ray optics for the Helmholtz equation, thus developping a systematic 
understanding with semiclassical methods. 
Imposing in a first step a mirror boundary with unit reflectivity makes the 
Helmholtz equation become identical to the Schr{\"{o}}dinger equation and in 
the short wavelength limit then Newtonian mechanics becomes applicable. 
The large shifts in resonance frequencies compared to the resonance spacings 
caused by the deformations prohibit the use of standard perturbation 
techniques as well~\cite{bowtie}. 

This paper therefore examines the propagation of rays like in a billiard, 
obeying the law of specular reflection at the boundary. 

With respect to the desired shape of a micro-disk, a circular disk, emitting 
light through frustrated total reflection is rather inefficient and 
isotropic~\cite{Phot-Bill}. This insight paved the way towards breaking the 
rotational symmetry. The \textit{flattened} quadrupole (oval) disks greatly 
improved emission and directionality, yet ``a further improvement may be 
expected if the remaining symmetries of reflection of the oval prototype are 
broken in an appropriate way.~\cite{Phot-Bill}''

I therefore suggest to consider a new type of basic deformation of circular 
disks: \eas deformations. As for a circle, radius and tangent always enclose 
an angle of \phtext. Increasing this angle by a constant angle $\delta$ 
deforms the circle to an equiangular spiral. 

In this paper I first describe this deformation in the language of 
two-di\-men\-sio\-nal geometric calculus~\cite{DH:CAGC}, which is equally 
suited to describing the classical hardwall shortwave limit~\cite{DH:NFII}, 
the application of semiclassical methods~\cite{lpath-eas} and the full quantum 
theory~\cite{DH:STA, CD:elec-ph}. This approach is further motivated by the 
fact that geometric calculus proves to be ``a unified language for 
mathematics and physics.~\cite{DH:uni-lang}''

After showing how geometric products of vectors describe rotations and 
reflections (see also~\cite{SG:inoreal}), I derive how the law of sinuses in 
the language of geometric calculus relates two successive reflections with 
each other. A detailed investigation of the consequences of this for 
classical ray propagation in equiangular spirals has already been carried out 
in~\cite{lpath-eas}. The present work only briefly discusses some new results 
on ray propagation and \eas geometry. It is then argued from a classical 
point of view that for the occurance of closed paths the gap 
(comp. figure~\ref{fg:of4}) should have unit reflectivity as well. Then 
two-dimensional geometric calculus is used for the general derivation of 
the constituting equations of closed paths in equiangular sprials. These 
constituting equations are first stated in terms of (bivector parts of) 
products of rotations (rotors) and radii of points of reflections. After 
that the explicite forms useful for numerical calculations are given. 

\bfg
	\begin{center}
	  \scalebox{1}{\includegraphics{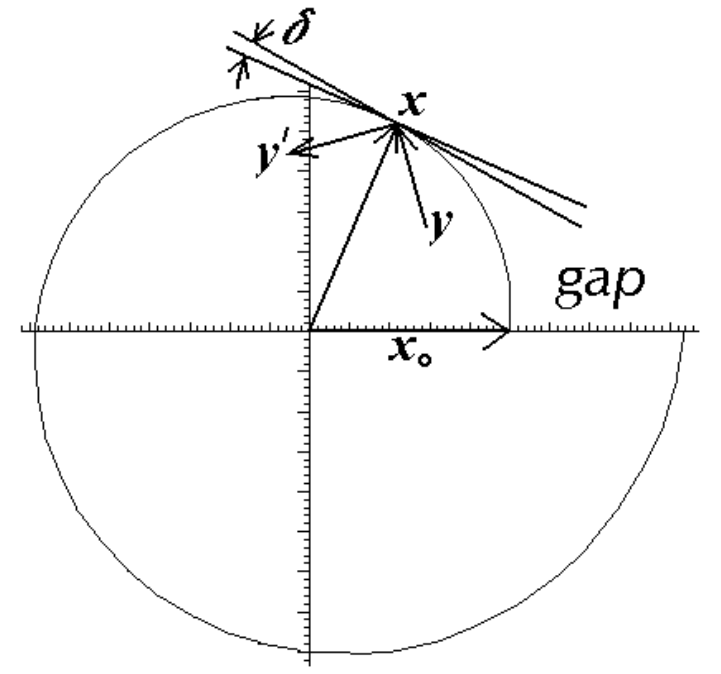}}
	\end{center}
  \protect\caption{An equiangular spiral with deformation parameter 
   $\delta = 0.1$. The tangent at \vect{x}~and a line perpendicular to 
   the radius vector \vect{x}~  are shown. They enclose the angle $\delta$.}
  \label{fg:of4}
\efg

The section on numerical results first briefly describes the numerical 
algorithm. This is followed by a detailed discussion of a set of basic 
closed paths with up to nine reflections on the \eas boundary plus one 
reflection at the gap. A major result is the general presence of degenerate 
modes with vertical reflections at the gap, whereas nondegenerate modes 
with angles of reflection less than \phtext~appear to exist only for 
very special values of the deformation parameter $\delta$.
Amongst the nondegenerate paths, the path with $n=3$ reflections on the \eas
boundary plus one reflection at the gap proves to be a 
\textit{stable asymmetric bow-tie} shaped light trajectory. Stability
is shown for a narrow subinterval of the $n=3$ nondegeneracy 
$\delta$-interval.

\section{Geometry of light paths in equiangular spiral disks}

In this section I will only summarize and slightly enhance the results 
of previous work~\cite{lpath-eas} without repeating most of the proves.

\subsection{Equiangular spiral described through geometric calculus}

The arena of geometry I will be dealing with is a \textit{real} two 
dimensional Euclidean vector space $\mathcal{E}_2$ representing a plane and 
its real geometric algebra $\mathcal{G}_2$. Fundamental for the notion of 
vector in geometric calculus is the associative geometric product of two 
vectors \vect{a}, \vect{b}:  
\bdm
 \vect{a} \vect{b} = 
 <\vect{a} \vect{b}> 
  + <\vect{a} \vect{b}>_2 =
  \vect{a} \cdot \vect{b} +
 \vect{a} \wedge \vect{b}
\edm
composed of the scalar inner product 
$<\vect{a} \vect{b}> = \vect{a} \cdot \vect{b}$ 
and the outer product 
$\vect{a} \wedge \vect{b}$. 
The latter simply represents the oriented area swept out by \vect{b}, as
it is displaced parallel along \vect{a}. It is also called a bivector 
$<\vect{a} \vect{b}>_2 = \vect{a} \wedge \vect{b}$ 
or in this case pseudoscalar, because its rank two is maximal in 
$\mathcal{G}_2$. 

The product of two vectors forms a spinor. With the help of the oriented unit 
area element ${\mathbf{i}} \in \mathcal{G}_2$ it can be written in 
exponential form  
\begin{eqnarray}
   \vect{a} \vect{b}
  & = &
     a \, b \, \, \exp( {\mathbf{i}}  \Phi )          \nonumber \\
  & = &
   a \, b \, \, (\cos \Phi + {\mathbf{i}} \sin \Phi)    
   \label{eq:gpexp}\\
  a & = & \sqrt{\vect{a} \vect{a}}  \nonumber
\end{eqnarray}
with ${\mathbf{i}}^2 = -1$.

For the special case that both \vect{a} and \vect{b} are unit vectors 
$(a=b=1)$ 
\be
  R_{ab} = \vect{a} \vect{b} = \exp( {\mathbf{i}}  \Phi )
  \label{eq:uurotor}
\ee
can be used to describe the rotation of \vect{a} into \vect{b}:
\be
  \vect{a} R_{ab} = \vect{b}.
  \label{eq:rotate}
\ee
This is the reason why $R_{ab}$ is called a rotor.

An equiangular spiral may now be described as
\be
  \vect{x} = \vind{x}{0} \exp({\mathbf{i}}  \Phi + t \Phi  )
  \label{eq:easdef}
\ee
with $t = \mbox{const.} \in \mathbf{R}$, $t > 0$, and $\Phi \in [0,2 \pi[$. 
The second term in the sum of the exponential describes the radial increase 
compared to a circle with radius $x_0$. 

It can be shown that the tangent $\partial_{\Phi} \vect{x}$ at any point 
\vect{x} has relative to the direction of the vector \vect{x} the constant 
angle 
\be
  \frac{\pi}{2} + \delta \,\,\, \mbox{ with } \,\,\, 
\delta = \arctan t = \tan^{-1} t. 
  \label{eq:tangle}
\ee

\subsection{Reflections inside an equiangular spiral}

A single reflection of a classical light ray as pictured in 
figure~\ref{fg:of4} propagating in the direction \vect{y} takes the 
simple product from 
\be
  \vect{y}^{\prime} = 
    R^{\dag}{}(-\delta) (-\unitv{x} \vect{y} \unitv{x}) R(-\delta)
  \stackrel{\mbox{2-dim.}}{=}
    -\unitv{x} \vect{y} \unitv{x} R(-2 \delta) 
  \label{eq:refl}
\ee
with $\unitv{x}=\vect{x}/x$, the rotor $R(-\delta)=\exp(-\mathbf{i} \delta)$, 
and its reverse (comp.~\cite{DH:CAGC}, p. 5) 
$R^{\dag}{}(-\delta) = \exp(\mathbf{i} \delta)$. 
The inner bracket $(-\unitv{x} \vect{y} \unitv{x})$
simply represents a reflection at a circle with radius vector \vect{x}, 
which due to Eq. (\ref{eq:tangle}) is followed by a clockwise rotation 
by the angle of $2 \delta$. The anticommutativity of the unit plane area 
element bivector \textbf{i} with any vector in its plane 
$\vect{x} \mathbf{i} = -\mathbf{i} \vect{x}$ explaines why 
$R^{\dag}{}(-\delta)$ becomes $R(-\delta)$ when shifted to the right as on 
the right side of the second equation in Eq. (\ref{eq:refl}). 

In this context it proves uesful to distinguish \textit{incidence from 
the right} for which the reflected ray leaves the reflecting equiangular 
spiral boundary to the left of the radius vector of the point of reflection, 
i.e., 
$\tp=\Theta + 2 \delta \in [0,\phtext + \delta[$ 
(comp. figure~\ref{fg:of4}) and \textit{incidence from the left} 
for which the reflected ray leaves to the right of the radius vector, 
i.e., $\tp \in [-\phtext + \delta,0[$.

\bfg
	\begin{center} 
	  {\scalebox{1}{\includegraphics
 {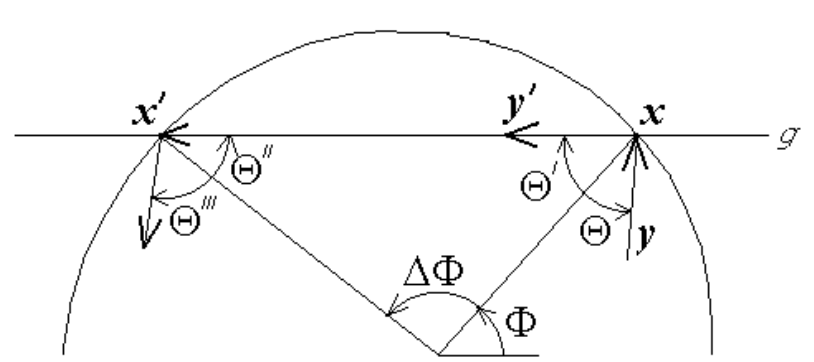}}}
	\end{center} 
  \caption{Two successive reflections at an \eas boundary forming the 
   triangle: origin, \vect{x}~and $\vect{x}^{\prime}$.}
  \label{fg:of5}
\efg

A typical succession of two reflections is shown in figure~\ref{fg:of5}. The 
origin and the two points of reflection \vect{x} and $\vect{x}^{\prime}$ 
form a triangle. Its three side vectors are related by 
\bdm
  \vect{x}^{\prime} = \vect{x} 
  + (\vect{x}^{\prime} - \vect{x}).
\edm
Hence 
\bea
  <(\vect{x}^{\prime} -\vect{x} ) \vect{x}^{\prime}>_2 
   =  <(\vect{x}^{\prime} - \vect{x})\vect{x}>_2, 
  \label{eq:sinlaw}
\eea
or equivalently $(\vect{x}^{\prime} - \vect{x}) \wedge \vect{x}^{\prime} 
   =  (\vect{x}^{\prime} - \vect{x}) \wedge \vect{x} $,
since the outer product of $\vect{x}^{\prime} - \vect{x}$ with itself 
necessarily vanishes. With the help of Eq. (\ref{eq:gpexp}) and by dividing 
Eq. (\ref{eq:sinlaw}) with $\mid \vect{x}^{\prime} - \vect{x} \mid $, and with 
\textbf{i} shows that Eq. (\ref{eq:sinlaw}) is nothing else but the familiar 
law of sinuses 
\be
 x^{\prime}{}\, \sin \tpp = x \, \sin (\pi - \tp) = x \, \sin \tp
  \label{eq:thgrow}
\ee
or simply
\bdm
  \sin \tp - \frac{x^{\prime}}{x} \sin \tpp = 0.
\edm
According to Eq. (\ref{eq:easdef}) the ratio of the two radii is 
$x^{\prime}/x = \exp[t(\Phi^{\prime}-\Phi)]$ and we therefore finally obtain
\be
  \sin \tp - \exp (t \DP) \sin \tpp = 0
  \label{eq:oneref}
\ee
with $\DP = \Phi^{\prime}-\Phi$.\footnote
{Equation (\ref{eq:oneref}) was derived in~\cite{lpath-eas} as well, 
yet the derivation presented here is much clearer and straightforward.} 

As the derivation above shows, Eq. (\ref{eq:oneref}) holds no matter whether 
the direction $\Phi = 0$ is not intersected by the line segment 
$\overline{\vect{x} \vect{x}^{\prime}}$ as in figure~\ref{fg:of5} or whether 
it is intersected.

For right incident rays Eq. (\ref{eq:oneref}) can be visualized as in 
figure~\ref{fg:of6}. The detailed examination of Eq. (\ref{eq:oneref}) yields 
that in general 
\bdm
  \tpp > \ta
\edm 
holds for right incident rays and 
\bdm
  \tpp < \ta
\edm
for left incident rays.

\bfg
	\begin{center} 
	  {\scalebox{1}{\includegraphics
    {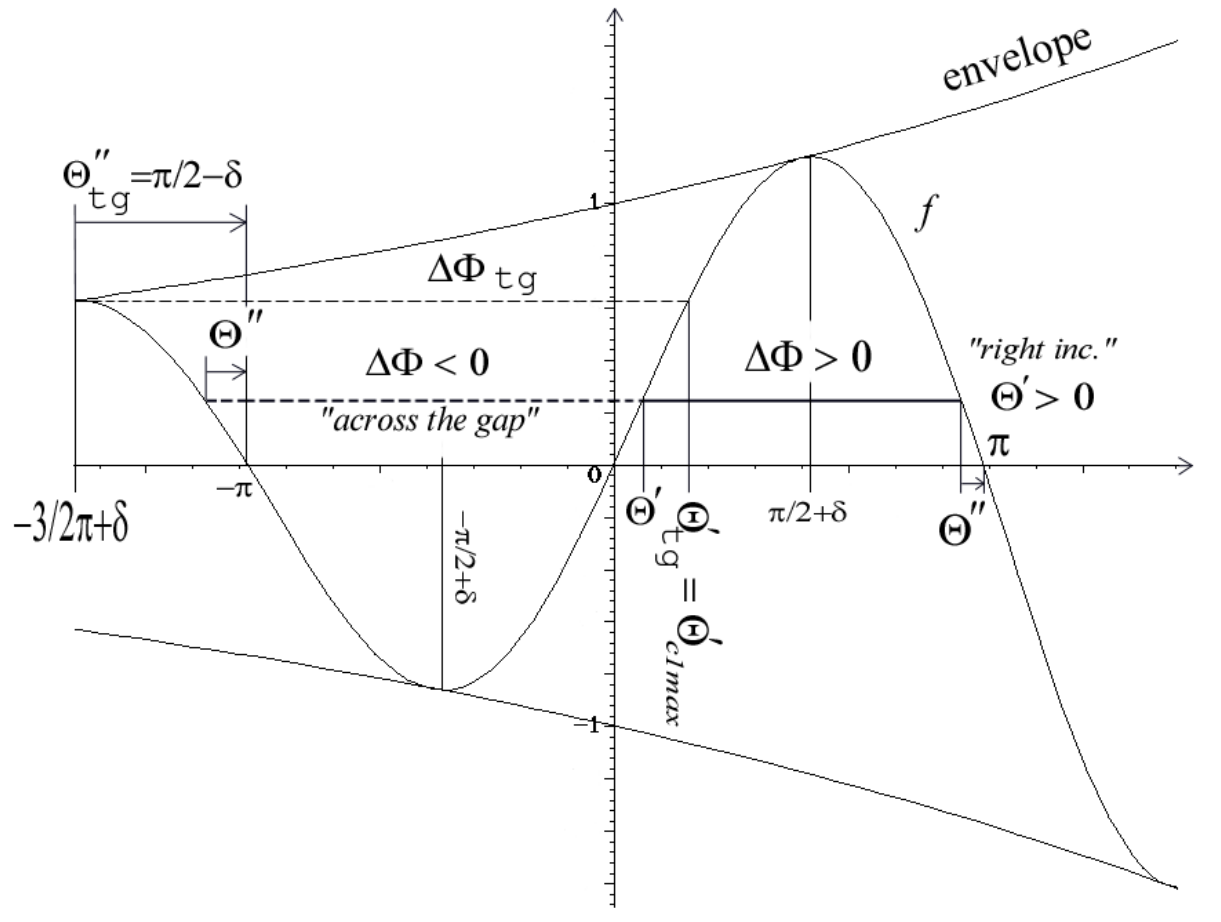}}}
	\end{center} 
  \caption{Graphical solution of Eq. (\ref{eq:oneref}) for successive 
   reflections of right incident rays. $\Delta \Phi_{tg}$, $\tp_{tg}$, 
   $\tp_{c1max}$ are all shown in figure~\ref{fg:of8}, and $f=\exp(t \tp) 
   \sin \tp$.}
  \label{fg:of6}
\efg

The general development will therefore be that right incident rays follow 
anticlockwise polygonal paths bending closer and closer to the boundary 
until they eventually escape. Left incident rays first start to perform 
clockwise polygonal motions yet bending further and further away from the 
boundary until they eventually change their \textit{state} into right 
incident rays with anticlockwise paths. An illustration corresponding to 
figure~\ref{fg:of6} for left incident rays can be found in~\cite{lpath-eas}, 
figure 11. 

As for rays leaving the equiangular spiral through the gap, 
figure~\ref{fg:of8} shows the range of angles under which this can happen: 
$\tp_{c1}{}(\Phi) < \tp < \tp_{c2}{}(\Phi)$ where $\Phi$ marks the point of 
the last reflection on the equiangular spiral boundary and the limiting angles 
correspond to incidence at $\mathbf{x}_0$ and $\mathbf{x}_{2\pi}$, 
respectively. An example of the nu\-me\-ri\-cal\-ly calculated dependencies of 
$\tp_{c1}{}(\Phi)$ and $\tp_{c2}{}(\Phi)$ from $\Phi$ is shown in the 
previously unpublished figure~\ref{fg:escape}. The actual angle and abscissa 
of escape as shown in figure~\ref{fg:of8} can be expressed as
\be
  \Phi_e = \Phi - \pi - \tp \,\, \mbox{and} \,\, 
  x_e = x \frac{\sin \tp}{\sin \Phi_e}.
  \label{eq:phxesc}
\ee

\bfg
	\begin{center} 
	  {\scalebox{1}{\includegraphics
    {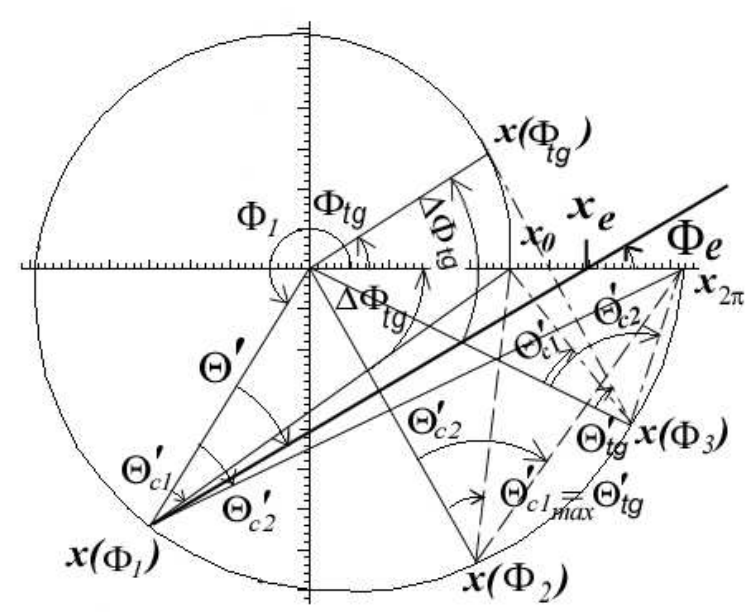}}}
	\end{center} 
  \caption{Angles $\tp_{c1}$ and $\tp_{c2}$ under which the gap, 
   i.e., \vind{x}{0}~and \vind{x}{2 \pi}~, respectively, are seen from 
   various boundary points $\vect{x}(\Phi_1)$, $\vect{x}(\Phi_2)$, 
   and $\vect{x}(\Phi_3)$ on the \eas. 
   $\overline{\vect{x}(\Phi_2)\vind{x}{0}}$ is part of the tangent 
   at \vind{x}{0}. $\tp_{tg}$ corresponds to striving tangential 
   incidence at the outside \eas boundary point $\vect{x}(\Phi_{tg})$.
   An escaping ray with angle of escape $\Phi_e$ and abscissa of escape 
   $x_e$ is shown [comp. Eq. (\ref{eq:phxesc})].} 
  \label{fg:of8}
\efg

\bfg
	\begin{center} 
	  {\scalebox{1}{\includegraphics
    {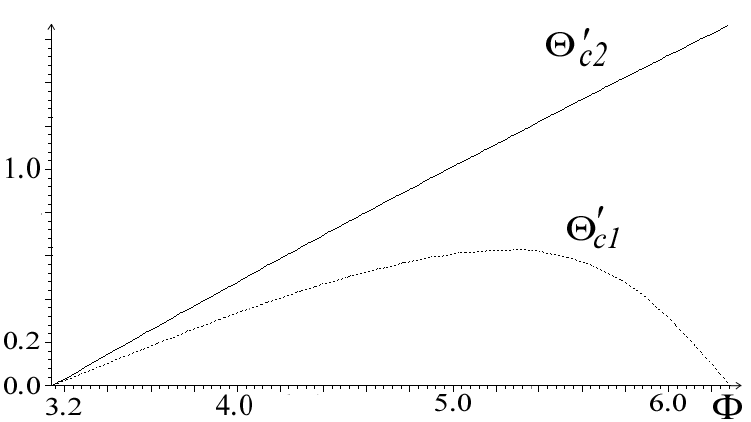}}}
	\end{center} 
  \caption{The abscissa $\Phi$ marks the point $\vect{x}(\Phi)$ of the 
   last reflection on the \eas boundary. $\tp_{c1}$ and $\tp_{c2}$ are 
   the angles under which both ends of the gap, i.e., \vind{x}{0}~and 
   \vind{x}{2 \pi}~, respectively, are seen from $\vect{x}(\Phi)$. 
   The two curves are calculated for $\delta = 0.1$.}
  \label{fg:escape}
\efg

One of the most interesting results about the geometry of light paths for 
equiangular spirals is that depending on where one locates a(n isotropic) 
source of light rays inside an equiangular spiral one may at the initial 
reflection either have only right incident rays for any ray emitted from the 
source or one may have both one angular sector with right incident rays and 
another with left incident rays. 

It was found that for sources located inside a certain area, described by a 
socalled critical equiangular spiral, only right incident rays will be 
emitted. This critical equiangular spiral has the same origin, and the same 
$t$ and $\delta$ as the original one, but is rotated anticlockwise by an 
angle of $\Phi_0 = \phtext -\delta$ and shrunk by a factor of 
$2 \sin \delta$. Outside this critical spiral both left and right handed 
angular sectors exist for the first reflection, delimited by the two tangents 
to the critical equiangular spiral which pass through the source of the light 
rays. 

An example  of a critical equiangular spiral is shown in figure~\ref{fg:of15} 
showing the sectors of right and left incidence for a source outside the 
critical equiangular spiral. 

\bfg
	\begin{center} 
	  {\scalebox{1}{\includegraphics
    {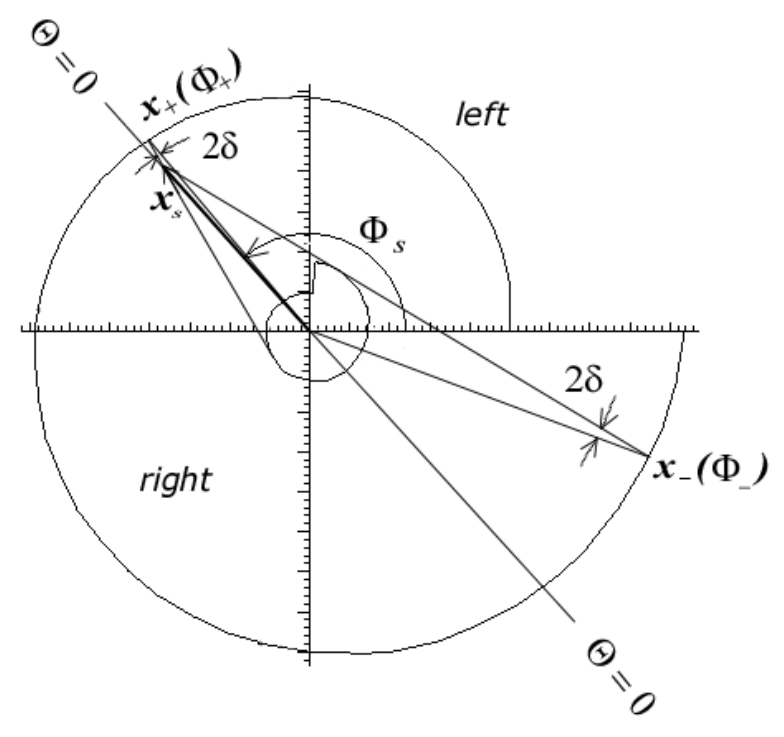}}}
	\end{center} 
  \caption{The inner \eas is the critical curve. For the source 
   \vind{x}{s}~outside it there exists one angular sector with 
   \textit{right} incidence for the first reflection and another sector 
   with \textit{left} incidence, respectively. These sectors are delimited 
   by the tangents to the critical equiangular spiral.}
  \label{fg:of15}
\efg

A further interesting geometrical property of this critical equiangular spiral 
is the fact that it can also be constructed as the \textit{envelope} to the 
set of once reflected rays, which originate at the origin. 

An instructive application of how easy it is to calculate arbitrary directed 
areas in the frame work of geometrical calculus is provided by the 
calculation of the oriented area of an equiangular spiral using the 
following line integral (comp.~\cite{DH:NFII}, p. 113):
\be
  A = \frac{1}{2} \int_0^{2\pi} \vect{x} \,\, \partial_{\Phi}{}\vect{x} d \Phi
  \label{eq:areaint}
\ee
where \vect{x} depends on $\Phi$ as given in Eq. (\ref{eq:easdef}). 
The derivative $\partial_{\Phi}\vect{x}$ is simply
\bdm
  \partial_{\Phi}\vect{x}=\vect{x}_0 \exp(\mathbf{i}\Phi + t \Phi)
(\mathbf{i}+t)
     = \vect{x} \mathbf{i} + \vect{x}t .
\edm
Inserting this back into (\ref{eq:areaint}) we obtain
\bea
  A & = & \frac{1}{2} \int_0^{2\pi} \vect{x} \wedge 
(\vect{x}\mathbf{i}) d \Phi 
  + \frac{1}{2}t \int_0^{2\pi} \underbrace{\vect{x} \wedge 
\vect{x}}_{=0} d \Phi
    \nonumber \\ & = & 
    \frac{1}{2} \mathbf{i} \int_0^{2\pi} x^2 d \Phi 
    = \frac{1}{2} \mathbf{i} \int_0^{2\pi} x^2_0 \exp(2 t \Phi) d \Phi
    \nonumber \\ & = & 
    \frac{1}{2} x^2_0 \frac{1}{2t} [ \exp(2 t \Phi) ]|^{2\pi}_0 
    =  \frac{x^2_0}{4t} [\exp(4 \pi t)-1] \mathbf{i}
    \nonumber \\ & = & 
    \frac{x^2_{2\pi}-x^2_0}{4t} \mathbf{i}
  \label{eq:easarea}
\eea
with $\vect{x}_{2\pi}=\vect{x}(\Phi=2\pi)$.

The relationship $\vect{x} \wedge (\vect{x}\mathbf{i})=x^2{}\mathbf{i}$
used here can be understood if one considers that $\vect{x}\mathbf{i}$ 
describes according to Eq. (\ref{eq:gpexp}) an anticlockwise rotation of 
\vect{x} by \phtext. \vect{x} and $\vect{x}\mathbf{i}$ are therefore 
perpendicular to each other implying that the scalar part 
$\vect{x} \cdot (\vect{x}\mathbf{i})$ of 
$\vect{x} (\vect{x}\mathbf{i})$ must vanish. The \textbf{i} in Eq. 
(\ref{eq:easarea}) is clearly recognized in its role as oriented unit 
area element.  

Taking into account that
\bea
  (\partial_{\Phi}{}\vect{x})^2
   & = & \vect{x}(\mathbf{i} + t) \vect{x} (\mathbf{i} + t)
   \nonumber \\
   & = & \vect{x} \vect{x} (-\mathbf{i} + t)  (\mathbf{i} + t)
   \nonumber \\
   & = & \vect{x}^2{} (1+t^2)
   \nonumber
\eea
we can also calculate the circumference of an equiangular spiral by another 
line integral
\bea
  U & = & \int_0^{2\pi} [(\partial_{\Phi}{}\vect{x})^2]^{1/2}{} d \Phi
      = (1+t^2)^{1/2}{} \int_0^{2\pi} x \, d \Phi
    \nonumber \\
    & = & (1+t^2)^{1/2}{} \,\, x_0 \int_0^{2\pi}\, \exp(t\Phi) dt 
    \nonumber \\
    & = & \left(\frac{1+t^2}{t^2}\right)^{1/2}{} x_0 [\exp(2\pi t)-1]
    \nonumber \\
    & \stackrel{(\ref{eq:tangle})}{=} &
      \frac{1}{\sin \delta} \,\,g
    \nonumber
\eea
where $g= \mid \vect{x}_{2 \pi} - \vect{x}_0 \mid = x_0 [\exp(2\pi t)-1]$, 
i.e., the length of the gap of the equiangular spiral as shown 
in~\ref{fg:of4}. 

Since the critical equiangular spiral is just a shrunk 
(by a factor of $2 \sin \delta$) 
version of the original one we can immediately 
infer its oriented area and circumference as
\bdm
   A_c  =  ( \sin \delta)^2{} \, A = \frac{1}{2} 
         \sin 2 \delta (x_{2\pi}^2-x_0^2) \mathbf{i} 
\edm
and
\bdm
   U_c  =  2 U \sin \delta = 2 g.
\edm
The surprising result is therefore that the circumference $U_c$ of the 
critical equiangular spiral measures exactly twice the length $g$ of the gap, 
i.e., the original spiral's straight line segment between \vind{x}{0} and 
\vind{x}{2\pi}.

This concludes the short discussion of some basic geometrical properties of 
equiangular spirals and light reflections inside it. 

The facts that eventually all light rays inside an equiangular spiral 
develop into right incident rays and that then the angels of reflection 
will increase continuously [comp. Eq. (\ref{eq:thgrow})] clearly 
demonstrates the impossibility to arrive at any sort of closed light path 
provided that the gap is kept ``\textit{open}''. Yet without closed paths an 
equiangular sprial disk could never qualify as a new disk laser geometry, 
like the by now familiar oval disks~\cite{Phot-Bill, Laser-Flash}. Possibly 
the simplest remedy for this might be to ``\textit{close}'' the gap by making 
it 100\% reflective as well. That this leads indeed to closed paths will be 
shown in the next section.

\section{Constituting equations for closed paths inside equiangular spirals}

An example of a closed path with a total of $n=9$ reflections at the 
equiangular spiral boundary plus one reflection at the gap is shown 
figure~\ref{fg:ex9}. In general the number of reflections at the \eas boundary 
may vary subject to certain restrictions, but as argued at the end of the 
last section, at least one reflection at the gap will always be present. 

\bfg
	\begin{center} 
	  {\scalebox{1}{\includegraphics
    {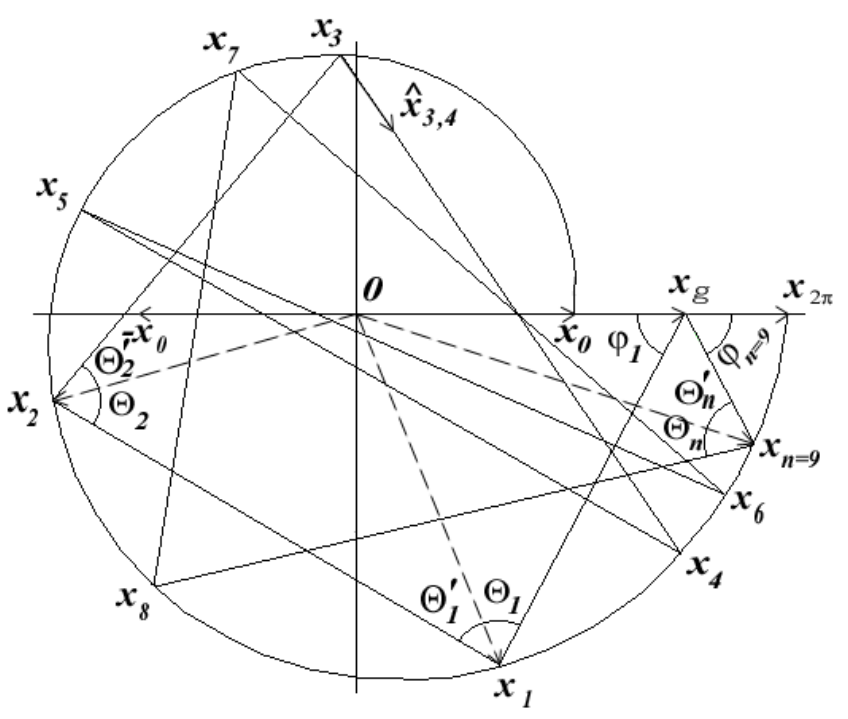}}}
	\end{center} 
  \caption{Closed path with $n=9$ (nondegenerate) reflections 
   $\vind{x}{1}(\Phi_1), \ldots , \vind{x}{n}(\Phi_n)$ along the 
   \eas boundary and one reflection at \vind{x}{g}~on the gap 
   $\overline{\vind{x}{0}\vind{x}{2\pi}}$.}
  \label{fg:ex9}
\efg

According to Eq. (\ref{eq:easdef}) a point of reflection on the \eas boundary 
is given by its position
\bea
  \vect{x}_k 
  & = & \vect{x}_0 \exp(\mathbf{i} \Phi_k + t \Phi_k) \nonumber \\
  & = & \unitv{x}_k x_k \label{eq:easpoint} \\
  & = & \unitv{x}_k x_0 \exp(t \Phi_k) \nonumber
\eea
where \uvind{x}{k} is the unit vector in the direction of \vind{x}{k} and 
$x_0 \exp(t \Phi_k)$ the scalar length of \vind{x}{k}. A useful notation 
in the following will be the vector of unit length \uvind{x}{k,k+1} in the 
direction of $\vind{x}{k+1} - \vind{x}{k}$, which is the path a light ray 
takes on the closed path from \vind{x}{k} to \vind{x}{k+1}:
\bdm
  \uvind{x}{k,k+1} = \frac{\vind{x}{k+1} - \vind{x}{k}}
                       {\mid \vind{x}{k+1} - \vind{x}{k} \mid}.
\edm
We already know from Eq. (\ref{eq:refl}) how \uvind{x}{k,k+1} and 
\uvind{x}{k-1,k} are related with each other by
\bea
  \uvind{x}{k,k+1} 
  & = & R^{\dag}{}(-\delta)(-\uvind{x}{k} \uvind{x}{k-1,k} \uvind{x}{k}) 
R(-\delta)
  \nonumber \\
   & \stackrel{\mbox{\small 2-dim.}}{=} &
   -\uvind{x}{k} \uvind{x}{k-1,k} \uvind{x}{k} {} R^2(-\delta).
  \label{eq:krefl}
\eea
Equation (\ref{eq:krefl}) can be used to successively expand each direction 
\uvind{x}{k,k+1} in terms of the ``first'' direction \uvind{x}{g,1} of the 
vector $\vind{x}{1} - \vind{x}{g}$, e.g.,
\be
  \uvind{x}{1,2} = -\uvind{x}{1} \uvind{x}{g,1} \uvind{x}{1} R(-2\delta)
  \label{eq:vxk12}
\ee
and
\bea
  \uvind{x}{2,3} & = & -\uvind{x}{2} \uvind{x}{1,2} \uvind{x}{2} R(-2\delta)
  \nonumber \\
  & = & 
  -\uvind{x}{2} 
  [-\uvind{x}{1} \uvind{x}{g,1} \uvind{x}{1} {} R(-2\delta)]\uvind{x}{2}
R(-2\delta)
   \label{eq:vxk23} \\
  & = &
  \uvind{x}{2}\uvind{x}{1}\uvind{x}{g,1}\uvind{x}{1}\uvind{x}{2}
  = R^{\dag}_{12} \uvind{x}{g,1} R_{12}
  \nonumber
\eea
with the rotor $R_{12}=\uvind{x}{1} \uvind{x}{2} 
= \exp(\mathbf{i}(\Phi_2 - \Phi_1))$
describing an anticlockwise rotation by $\DP_{12}=\Phi_2 - \Phi_1$.
It is to be noted that the two rotors $R(-2\delta)$ 
cancelled each other because 
$R(-2\delta)\uvind{x}{2} = \uvind{x}{2}R(2\delta)$, and 
$R(2\delta)R(-2\delta) = R(2\delta-2\delta) = R(0) = \exp(0 \mathbf{i}) = 1$.
The next direction along the closed light path can be expanded as
\bea
  \uvind{x}{3,4} 
  & = & -\uvind{x}{3} \uvind{x}{2,3} \uvind{x}{3} R(-2\delta) 
  \nonumber \\
  & = & -\uvind{x}{3} R^{\dag}_{12}{} \uvind{x}{g,1}R_{12}{} 
\uvind{x}{3} R(-2\delta).
  \label{eq:vxk34}
\eea
The above expressions for \uvind{x}{1,2}, \uvind{x}{2,3}, and \uvind{x}{3,4} 
given in (\ref{eq:vxk12}),(\ref{eq:vxk23}), and (\ref{eq:vxk34}), 
respectively, are sufficient in order to understand the construction of
\bea
  \uvind{x}{k,k+1} 
  & = &
  R^{\dag}_{k-1k}{}R^{\dag}_{k-3k-2}{} \ldots R^{\dag}_{12}{} 
  \uvind{x}{g,1}  R_{12}
  \nonumber \\
  && 
  \ldots R_{k-3k-2}R_{k-1k}
  \label{eq:keven}
\eea
for $k=2l, \mbox{ and } l\in \mathbf{N}$; and
\bea
  \uvind{x}{k,k+1} & = &
  -\uvind{x}{k}R^{\dag}_{k-2k-1}{}R^{\dag}_{k-4k-3}{} 
  \ldots R^{\dag}_{12}{} \uvind{x}{g,1}  R_{12}
  \nonumber \\
  && 
  \ldots R_{k-4k-3}R_{k-2k-1} \uvind{x}{k} R(-2\delta)
  \label{eq:kodd}
\eea
for $k=2l+1, \mbox{ and } l\in \mathbf{N}$.
In the very same way as in Eq. (\ref{eq:sinlaw}) we have for each triangle 
formed by the origin and two successive points of reflection \vind{x}{k} 
and 
\vind{x}{k+1} on the \eas
\be
  <\uvind{x}{k,k+1}\vind{x}{k}>_2 =<\uvind{x}{k,k+1}\vind{x}{k+1}>_2
  \label{eq:conbase}
\ee

Utilizing the expressions (\ref{eq:keven}) and (\ref{eq:kodd}) we can proceed 
now to derive the first set of $n-1$ constituting equations necessary for a 
closed path. In the following I will first demonstrate the steps involved for 
$k=1,2$, and 3. With this experience in hand it will then be easy to 
understand the derivation of the general form for any 
$k\in \mathbf{N}, k<n$. After that I will show how the closing condition 
yields the last $n$th constituting equation. 

Let us therefore begin with Eq. (\ref{eq:conbase}) for $k=1$:
\bdm
  <\uvind{x}{1,2}\vind{x}{1}>_2 =<\uvind{x}{1,2}\vind{x}{2}>_2
\edm
Inserting \vind{x}{1}, \vind{x}{2} and \uvind{x}{1,2} according to Eqs. 
(\ref{eq:easpoint}) and (\ref{eq:vxk12}) we obtain
\bea
  <\uvind{x}{1}\uvind{x}{g,1}\uvind{x}{1}R(-2\delta)\uvind{x}{1}x_{1}>_2 
  \hspace*{1cm}
   && \nonumber \\ 
  = <\uvind{x}{1}\uvind{x}{g,1}\uvind{x}{1}R(-2\delta)\uvind{x}{2}x_{2}>_2.
  && \nonumber
\eea
Interchanging the rotor $R(-2\delta)$ with \uvind{x}{1} and \uvind{x}{2}, 
respectively, and dividing by $x_1$ we get
\be
  <\uvind{x}{1}\uvind{x}{g,1}R(2\delta)>_2 
   = \frac{x_2}{x_1}<\uvind{x}{1}\uvind{x}{g,1}\uvind{x}{1}\uvind{x}{2}
R(2\delta)>_2
  \label{eq:eq1pre}
\ee
where we have used the fact that \uvind{x}{1} is a unit vector, i.e., 
$\uvind{x}{1}^2 = 1$. As seen from figure~\ref{fg:ex9} \uvind{x}{1} and 
\uvind{x}{g,1} enclose the angle $\ta_1$ which 
(because of the \textit{left} incidence at \vind{x}{1}) 
will always be negative:
\be
  \uvind{x}{1}\uvind{x}{g,1} = R(\ta_1).
  \label{eq:t1rotor}
\ee 
Equation (\ref{eq:eq1pre}) can therefore be rewritten as
\bea
  <R(\ta_1)R(2\delta)>_2 
  & = & <R(\ta_1+ 2\delta)>_2 
  \nonumber \\
  & \hspace*{-2cm} = & \hspace*{-1cm} \frac{x_2}{x_1}<R(\ta_1+ 2\delta)
R_{12}>_2
  \label{eq:eq1}
\eea
since $R(\beta_1)R(\beta_2) = R(\beta_1+\beta_2) = R(\beta_2)R(\beta_1)$ 
(additivity of the angles and commutativity of rotations in the same plane 
with a common center of rotation). 

Towards the end of the current section I will proceed to replace the rotors 
$R$ and the amplitudes $x_k$ by their respective exponential expressions. 
The extraction of the bivector parts $<>_2$ as in Eq. (\ref{eq:gpexp}) will 
then give the final somewhat unwieldy explicitely transcendental form of 
the constituting equations necessary for their concrete numerical evaluation. 

Next we take Eq. (\ref{eq:conbase}) for $k=2$:
\bdm
  <\uvind{x}{2,3}\vind{x}{2}>_2 =<\uvind{x}{2,3}\vind{x}{3}>_2
\edm
Inserting \vind{x}{2}, \vind{x}{3} and \uvind{x}{2,3} according to Eqs. 
(\ref{eq:easpoint}) and (\ref{eq:vxk23}) gives
\be
  <R^{\dag}_{12}\uvind{x}{g,1}R_{12}\uvind{x}{2}>_2 =\frac{x_3}{x_2} 
<R^{\dag}_{12}\uvind{x}{g,1}R_{12}\uvind{x}{3}>_2.
  \label{eq:eq2pre}
\ee
The only difference to the calculation for $k=1$ is that we need to first 
express \uvind{x}{3} in terms of \uvind{x}{1} using Eq. (\ref{eq:rotate}):
\be
  \uvind{x}{3} = \uvind{x}{1} R_{13}. 
  \label{eq:1r3}
\ee
With the help of Eq. (\ref{eq:1r3}) and taking into account that 
$R_{12}\uvind{x}{2} = \uvind{x}{2}R^{\dag}_{12} = \uvind{x}{2}R_{21}= 
\uvind{x}{1}$, and that $R_{12} \uvind{x}{3} = \uvind{x}{3} R^{\dag}_{12}$, 
Eq. (\ref{eq:eq2pre}) attaines the form
\be
  <R^{\dag}_{12}\uvind{x}{g,1}\uvind{x}{1}>_2 =\frac{x_3}{x_2}
<R^{\dag}_{12}\uvind{x}{g,1}\uvind{x}{1}R_{13}R^{\dag}_{12}>_2.
  \label{eq:eq2pre1}
\ee
The product of the rotors $R_{13}R^{\dag}_{12}$ means first to rotate a 
vector anticlockwise by $\Phi_3-\Phi_1$ and then to rotate it back 
clockwise by $\Phi_2-\Phi_1$:
\be
  R_{13}R^{\dag}_{12} = \uvind{x}{1}\uvind{x}{3}\uvind{x}{2}\uvind{x}{1}
  = \uvind{x}{1}R_{32}\uvind{x}{1} = R^{\dag}_{32} = R_{23}.
  \label{eq:r23}
\ee
Equation (\ref{eq:eq2pre1}) together with the relations (\ref{eq:t1rotor}) and 
(\ref{eq:r23}) results therefore in
\be
  <R(\ta_1)R_{12}>_2 = \frac{x_3}{x_2}<R(\ta_1)R_{12}R^{\dag}_{23}>_2
  \label{eq:eq2}
\ee
where I have already applied a final reversion to both sides of the equation. 

Setting $k=3$ for Eq. (\ref{eq:conbase}) gives
\bdm
  <\uvind{x}{3,4}\vind{x}{3}>_2 =<\uvind{x}{3,4}\vind{x}{4}>_2.
\edm
By inserting \vind{x}{3}, \vind{x}{4} and \uvind{x}{3,4} according to
(\ref{eq:easpoint}) and (\ref{eq:vxk34}) we obtain
\bea
  <\uvind{x}{3}R^{\dag}_{12}\uvind{x}{g,1}R_{12}\uvind{x}{3}
  R(-2\delta)\uvind{x}{3}>_2  
  \hspace{2cm} 
  && \nonumber \\
  = \frac{x_4}{x_3}<\uvind{x}{3}R^{\dag}_{12}\uvind{x}{g,1}
    R_{12}\uvind{x}{3}R(-2\delta)\uvind{x}{4}>_2
  \label{eq:eq3pre}
\eea
Manipulating Eq. (\ref{eq:eq3pre}) according to the by now familiar rules 
of geometric calculus we arrive at
\bea
  <R(\ta_1+2\delta)R_{12}R^{\dag}_{23}>_2 \hspace{2cm} && \nonumber \\
  =  \frac{x_4}{x_3}<R(\ta_1+2\delta)R_{12}R^{\dag}_{23}R_{34}>_2 
  \label{eq:eq3}
\eea

The explicite derivations of the first three constituting Eqs. 
(\ref{eq:eq1}), (\ref{eq:eq2}) and (\ref{eq:eq3}) are in my view sufficient 
for understanding
the general result for arbitrary $k < n$:
\bea
  &&\hspace{-1cm}<R(\ta_1+2\delta)R_{12}R^{\dag}_{23}\ldots R_{k-2k-1}
R^{\dag}_{k-1k}>_2 
  \nonumber \\
  &=& \frac{x_{k+1}}{x_k}<R(\ta_1+2\delta)R_{12} R^{\dag}_{23} 
  \nonumber \\
  && \ldots R^{\dag}_{k-1k}R_{kk+1}>_2
  \label{eq:eqkodd}
\eea
for $k=2l+1, \mbox{ and } l\in \mathbf{N}$; and
\bea
  &&\hspace{-1cm}<R(\ta_1)R_{12}R^{\dag}_{23}
  \ldots R^{\dag}_{k-2k-1}R_{k-1k}>_2 
  \nonumber \\
  &=& \frac{x_{k+1}}{x_k}<R(\ta_1)R_{12} R^{\dag}_{23} 
  \nonumber \\
  && \ldots R_{k-1k}R^{\dag}_{kk+1}>_2
  \label{eq:eqkeven}
\eea
for $k=2l, \mbox{ and } l\in \mathbf{N}$.

With Eqs. (\ref{eq:eqkodd}) and (\ref{eq:eqkeven}) we have already found all 
$n-1$ constituting equations which describe the light path between the $n$ 
points of consecutive reflections along the \eas boundary. Now we must take 
the last reflection at the gap into account in order to really close the path 
back into itself. 
Looking at figure~\ref{fg:ex9} we see that hence closing means 
$\varphi_1 = \varphi_n$. In order to express this properly one may, e.g., use 
two triangles, each containing one of these angles or its $\pi - \varphi$ 
\label{pg:pep}complement: the ones that shall be used here are formed by 
the origin, \vind{x}{g} and \vind{x}{1} or \vind{x}{n}, respectively. 
Applying the law of sinuses (\ref{eq:sinlaw}) to both triangles we get:
\bea
  <\vind{x}{1}\uvind{x}{g,1}>_2 
  & \stackrel{(\ref{eq:sinlaw})}{=} &
  <\vind{x}{g}\uvind{x}{g,1}>_2 
  \nonumber \\
  & = & x_g <\uvind{x}{0}\uvind{x}{g,1}>_2 
  \nonumber \\
  & = & - x_g \mathbf{i} \sin \varphi_1 
  \nonumber 
\eea
and
\bea
  <\vind{x}{n}\uvind{x}{n,g}>_2 
  & \stackrel{(\ref{eq:sinlaw})}{=} &
  <\vind{x}{g}\uvind{x}{n,g}>_2 
  \nonumber \\
  & = & x_g <\uvind{x}{0}\uvind{x}{n,g}>_2 
  \nonumber \\
  & = & x_g \mathbf{i} \sin \varphi_n.
  \nonumber 
\eea
Hence the equality $\varphi_1 = \varphi_n$ has as a consequence
\be
  -<\vind{x}{1}\uvind{x}{g,1}>_2 = <\vind{x}{n}\uvind{x}{n,g}>_2 .
  \label{eq:eqnpre}
\ee
Now \uvind{x}{n,g} is the direction of the light ray reflected at \vind{x}{n}, 
i.e., that according to Eq. (\ref{eq:krefl})
\be
  \uvind{x}{n,g} = -\uvind{x}{n} \uvind{x}{n-1,n} \uvind{x}{n} R(-2\delta).
  \label{eq:refln}
\ee
But \uvind{x}{n-1,n} is already known to us from Eqs. (\ref{eq:keven}) or 
(\ref{eq:kodd}), respectively. For the case that $n-1$ is even we can rewrite 
Eq. (\ref{eq:eqnpre}) therefore with the help of (\ref{eq:refln}) and 
(\ref{eq:keven}) as
\bea
  && \hspace{-1cm} -<\uvind{x}{1}\uvind{x}{g,1}>_2 
  \nonumber \\
  & = & 
  \frac{x_n}{x_1}<\uvind{x}{n}(-\uvind{x}{n}) 
                  R^{\dag}_{n-2n-1} \ldots R^{\dag}_{12}
                  \uvind{x}{g,1} 
                  R_{12} 
  \nonumber \\
  &&
  \ldots R_{n-2n-1}\uvind{x}{n}R(-2\delta)>_2
  \nonumber \\
  & = & 
  -\frac{x_n}{x_1}<R^{\dag}_{n-2n-1} \ldots R^{\dag}_{12}
                  \uvind{x}{g,1} 
                  R_{12} 
  \nonumber \\
  &&
  \ldots R_{n-2n-1} \uvind{x}{1}R_{1n}R(-2\delta)>_2
  \nonumber \\
  & = & 
  -\frac{x_n}{x_1}<R^{\dag}_{n-2n-1} \ldots R^{\dag}_{12}
                  \uvind{x}{g,1} \uvind{x}{1}
                  R^{\dag}_{12} 
  \nonumber \\
  &&
  \ldots R^{\dag}_{n-2n-1}R_{1n}R(-2\delta)>_2
  \nonumber \\
  & = & 
  -\frac{x_n}{x_1}<R^{\dag}_{n-2n-1} \ldots R^{\dag}_{12}
                  R^{\dag}(\ta_1)
                  R_{23} 
  \nonumber \\
  &&
  \ldots R_{n-1n}R^{\dag}(-2\delta)>_2
  \nonumber
\eea
where we have used that $\uvind{x}{n} = \uvind{x}{1}R_{1n}$, that 
$\uvind{x}{g,1}\uvind{x}{1}= (\uvind{x}{1}\uvind{x}{g,1})^{\dag} 
= R^{\dag}(\ta_1)$,
and that 
$R^{\dag}_{12} \ldots R^{\dag}_{n-2n-1}R_{1n} = R_{23} \ldots R_{n-1n}$.
Reversing the right side, which swallows up the minus sign and rearranging 
the rotors we end up with
\bea
  -<R(\ta_1)>_2 & = & \frac{x_n}{x_1}<R(\ta_1+2\delta)
  R_{12}R^{\dag}_{23}R_{34} 
  \nonumber \\
  &&
  \ldots R_{n-2n-1}R^{\dag}_{n-1n}>_2.
  \label{eq:eqnodd}
\eea

In the same way, just using Eq. (\ref{eq:kodd}) instead of Eq. (\ref{eq:keven})
we obtain for odd values of $n-1$:
\bea
  -<R(\ta_1)>_2 & = & \frac{x_n}{x_1}<R(\ta_1)R_{12}R^{\dag}_{23}R_{34} 
  \nonumber \\
  &&
  \ldots R^{\dag}_{n-2n-1}R_{n-1n}>_2.
  \label{eq:eqneven}
\eea

We may now summarize: for $n$ reflections along the \eas boundary plus one 
reflection at the gap we have now established all $n$ con\-sti\-tu\-ting 
equations. That is, $n-1$ constituting equations for $k=1, \ldots ,n-1$ of 
the form (\ref{eq:eqkodd}) and (\ref{eq:eqkeven}) for the odd and even values 
of $k$, respectively, plus one constituting equation of the form 
(\ref{eq:eqnodd}) or (\ref{eq:eqneven}) for $n$ being odd or even, 
respectively. 

In order to finally see what such a system of $n$ constituting equations 
looks like in explicitely transcendental form which is necessary for the 
numerical treatment in the next section, I will give here the constituting 
equations for the example of figure~\ref{fg:ex9} with $n=9$. 
This explicite form is obtained by in\-ser\-ting the rotors $R$ according to 
(\ref{eq:uurotor}) and the amplitudes $x_k$ according to 
(\ref{eq:easdef}) and (\ref{eq:easpoint}), respectively, into Eqs. 
(\ref{eq:eqkodd}), (\ref{eq:eqkeven}) and (\ref{eq:eqnodd}), and by 
extracting the $<>_2$ bivector parts. 
In order to simplify the notation I define $\Delta_{ij} = \Phi_i - \Phi_j$ 
with $1 \leq i,j \leq n$. 
\bea
  &&\hspace{-.9cm}
  \sin(\tao+2\delta)
  \nonumber \\ &=&
  \exp(-t\Delta_{12}) \sin(\tao+2\delta-\Delta_{12})
  \nonumber \\ &&\hspace{-.9cm}
  \sin(\tao-\Delta_{12})
  \nonumber \\ &=&
  \exp(-t\Delta_{23}) \sin(\tao-\Delta_{12}+\Delta_{23})
  \nonumber \\ &&\hspace{-.9cm}
  \sin(\tao+2\delta-\Delta_{12}+\Delta_{23})
  \nonumber \\ &=&
  \exp(-t\Delta_{34}) \sin(\tao+2\delta-\Delta_{12}+\Delta_{23}-\Delta_{34})
  \nonumber \\ &&\hspace{-.9cm}
  \sin(\tao-\Delta_{12}+\Delta_{23}-\Delta_{34})
  \nonumber \\ &=&
  \exp(-t\Delta_{45}) 
\sin(\tao-\Delta_{12}+\Delta_{23}-\Delta_{34}+\Delta_{45})
  \nonumber \\ &&
  \ldots
  \nonumber \\ &&\hspace{-.9cm}
  \sin(\tao-\Delta_{12}+\Delta_{23}-\Delta_{34}+\Delta_{45}-\Delta_{56}
  \nonumber \\
  &&
  +\Delta_{67}-\Delta_{78})
  \nonumber \\ &=&
  \exp(-t\Delta_{89}) \sin(\tao-\Delta_{12}+\Delta_{23}-\Delta_{34}+\Delta_{45}
  \nonumber \\
  &&
  -\Delta_{56}+\Delta_{67}-\Delta_{78}+\Delta_{89})
  \nonumber \\ &&\hspace{-.9cm}
  -\sin\tao
  \nonumber \\ &=&
  \exp[-t(\Delta_{12}+\Delta_{23}+\Delta_{34}+ \ldots +\Delta_{89})]
  \nonumber \\ &&
  \sin(\tao+2\delta-\Delta_{12}+\Delta_{23}-\Delta_{34}+ \ldots 
  \nonumber \\
  &&
  -\Delta_{78}+\Delta_{89})
  \nonumber
\eea
The system of $n$ constituting equations determines the $n$ variables 
$\Delta_{kk+1} (k=1,\ldots,n-1)$, and \tao~completely. But as for concrete 
applications one is of course interested to calculate the angles $\Phi_k 
(k=1,\ldots,n)$ itself and not only their differences. I will therefore derive 
a relation giving the explicite dependence of $\Phi_1$ with respect to 
$\Delta_{kk+1} 
(k=1,\ldots,n-1)$, and \tao.

We start with the relationship $\varphi_1 = \varphi_n$ already referred to on 
page~\pageref{pg:pep} for the derivation of the $n$th constituting equation. 
It is equivalent to
\bdm
  \exp(\mathbf{i}\varphi_1) = \exp(\mathbf{i}\varphi_n),
\edm 
which may in turn be equivalently expressed according to (\ref{eq:uurotor}) 
by the full geometric products of pairs of unit vectors enclosing these 
angles, i.e
\be
  (-\uvind{x}{0})\uvind{x}{g,1} = \uvind{x}{g,n}\uvind{x}{0}.
  \label{eq:prodpep}
\ee
It was the bivector part of this relationship that led to Eq. 
(\ref{eq:eqnpre}). 

For the case of $n-1$ being even, Eq. (\ref{eq:keven}) allows us to replace 
\uvind{x}{g,1} on the left side of (\ref{eq:prodpep}) by
\be
  \uvind{x}{g,1} = R_{12}R_{34} \ldots R_{n-2n-1} \uvind{x}{n-1,n} 
R^{\dag}_{n-2n-1}\ldots R^{\dag}_{34} R^{\dag}_{12}.
  \label{eq:uxg1}
\ee
using Eq. (\ref{eq:refl}) the reflection of \uvind{x}{n-1,n} at \vind{x}{n} 
into $\uvind{x}{n,g} = -\uvind{x}{g,n} $ is described by
\be 
  -\uvind{x}{g,n} = -\uvind{x}{n}\uvind{x}{n-1,n}\uvind{x}{n} 
R(-2\delta)
  \label{eq:reflatn}
\ee
Inserting (\ref{eq:uxg1}) and (\ref{eq:reflatn}) back into 
Eq. (\ref{eq:prodpep}) and multiplying with $-\uvind{x}{0}$ from the 
left yields
\bea
  &&\hspace{-1cm} 
  R_{12}R_{34} \ldots R_{n-2n-1} \uvind{x}{n-1,n} R^{\dag}_{n-2n-1}
\ldots R^{\dag}_{34} R^{\dag}_{12}
  \nonumber \\
  &=&
  -\uvind{x}{0}\uvind{x}{n}\uvind{x}{n-1,n}\uvind{x}{n}R(-2\delta)\uvind{x}{0}
  \label{eq:phi1pre}\\
  & = & -R_{0n}\uvind{x}{n-1,n}R^{\dag}_{0n}R(2\delta).
  \nonumber
\eea
Reshuffling all rotors to the left of each side of Eq.~(\ref{eq:phi1pre}), 
multiplying with \uvind{x}{n-1,n} from the right and expressing the negative 
sign in exponential form as well results in
\bdm
  R^2_{12}R^2_{34} \ldots R^2_{n-2n-1} = \exp(\mathbf{i}\pi)R^2_{0n}
R^2(-\delta)
\edm
or after taking the squareroot in
\be
  R_{0n} = \pm \exp(\mathbf{i}\frac{\pi}{2}) R(\delta)R_{12}R_{34} 
\ldots R_{n-2n-1}.
  \label{eq:phi1pre1}
\ee
Equating the exponentials of (\ref{eq:phi1pre1}) up to multiples of 
$2\pi$ and considering the sign ambiguity we get
\bdm
  \Phi_n = (2m+1) \ph + \delta - \Delta_{12}- \Delta_{34} \ldots 
            - \Delta_{n-1n-2}, m\in\mathbf{Z}
\edm
since 
\bdm
  R_{0n} = \uvind{x}{0}\uvind{x}{n}= \exp{\mathbf{i} \Phi_n}.
\edm
And since $\Phi_1 = \Phi_n + \Delta_{1n} = \Phi_n 
           + \sum_{k=1}^{n-1} \Delta_{kk+1}$ 
we finally obtain 
\bdm
  \Phi_1 = (2m+1) \ph + \delta + \Delta_{23} + \Delta_{45} + \ldots 
           + \Delta_{n-1n}, m\in\mathbf{Z}.
\edm
A similar calculation for $n-1$ being odd results in the corresponding 
expression
\bdm
  \Phi_1 = m \pi - \tao + \Delta_{23} + \Delta_{45} + \ldots 
           + \Delta_{n-1n}, m\in\mathbf{Z}.
\edm
It is obvious from figure~\ref{fg:ex9} that the $m$ 
in the above expressions for 
$\Phi_1$ must be chosen such that $\pi<\Phi_1<2\pi$.

\section{Numerical results}

\subsection{Remarks about the numerical technique}

In order to evaluate the $n$ constituting equations derived in the last 
section short numercial algorithms were written with MAPLE V using the command 
\textit{fsolve}. The highly transcendental character of the equations made 
it necessary to first ``guess'' very narrow intervals (e.g., $\pm 10^{-5}$) 
around one point of the actual solution. In order to do this a light ray 
tracing algorithm was developped. By applying it repeatedly to a range of 
initial conditions 
$({x}_{g}, \varphi_1)$ the resulting 
$({x}_{g}^{\prime}, \varphi_n)$ after tracing the rays once along a closed 
path were obtained and the errors 
$({x}_{g}^{\prime}-{x}_{g}, \varphi_n-\varphi_1)$ 
were plotted. Gradually minimizing these errors yielded the ``guessed'' 
range intervals for the application of MAPLE V`s \textit{fsolve} with 
sufficient accuracy. 

The curves $x_g(\delta)$ and $\varphi_1(\delta)$ were then calculated by first 
finding the exact point of solution (only limited by the digital accuracy 
chosen for the application of MAPLE V) in the range interval and successive 
small stepwise variations of $\delta$ starting from the first exact point of 
solution. 

In practice it was much easier to obtain \textit{degenerate} paths with 
$\varphi_1 = \varphi_n = \ph$ than \textit{nondegenerate} paths with 
$ 0 < \varphi_1 < \ph$.

\subsection{Degenerate closed paths with $\varphi_1 = \varphi_n = \ph$}

It is obvious that the simplest closed light path consists of just one 
vertical line\footnote{This may correspond to a stable, curved mirror 
Fabry-Perot resonator~\cite{bowtie}.}
 as indicated in the inset of figure~\ref{fg:d1}. As we know, e.g., from 
Eq. (\ref{eq:refl}), the angle of incidence \ta~ and the angle of the 
reflected ray \tp~ relative to the radius vector of the point of reflection 
are related by
\bdm
  \tp = \ta + 2 \delta
\edm
In this case we have $\ta < 0$ and $\abs{\tp} = \abs{\ta}$. Hence
\bdm
  \tp = - \ta = \delta.
\edm 
The location of the point of reflection on the \eas boundary will obviously be
\bdm
  \Phi = \frac{3}{2} \pi + \delta
\edm 
and as shown in figure~\ref{fg:d1} the abscissa of reflection $x_g$ located on 
the gap will be
\bdm
  x_g = x_0 \exp (t \Phi) \sin \delta
\edm
under the condition that $x_g \geq x_0$. This will be the case for
\bdm
  \delta > \delta_c \approx 0.2638158642...\hspace{1cm}.
\edm
$x_g$ will become infinite for $\delta \rightarrow \ph$. 

\bfg
	\begin{center} 
	  {\scalebox{1}{\includegraphics
    {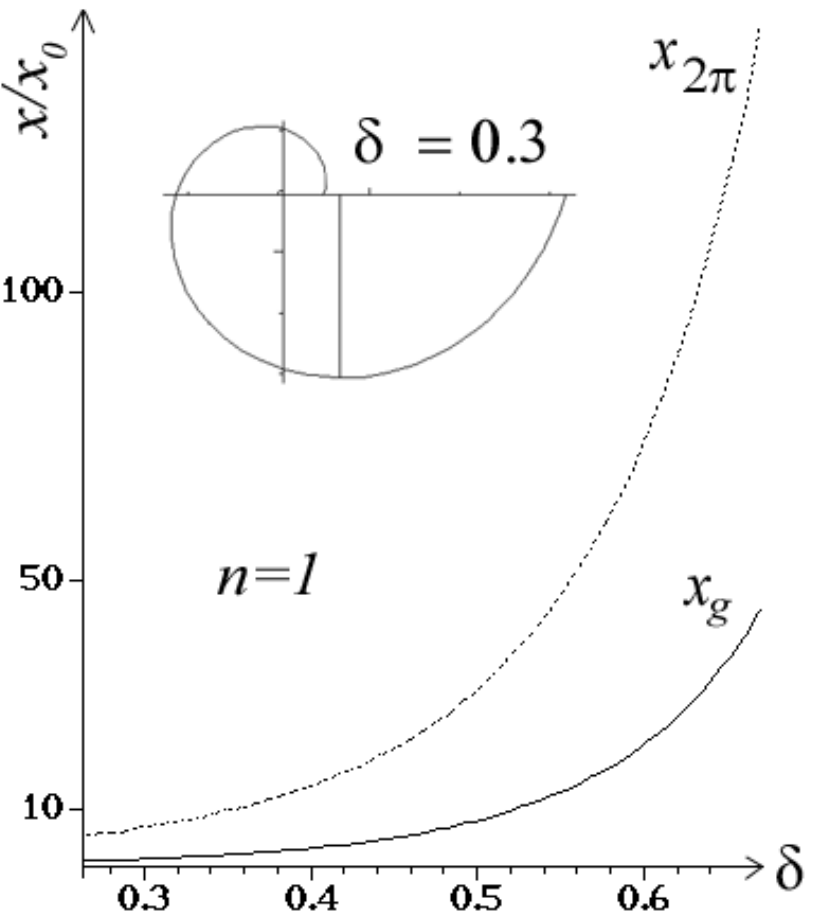}}}
	\end{center} 
  \caption{$x_{2\pi}$ and the point of reflection $x_{g}$ on the gap 
   for the degenerate closed path with $n=1$ reflections on 
   the \eas boundary. 
   An example of such a closed path is shown in the inset for $\delta=0.3$. 
   The abscissa $\delta$ actually extends up to $\delta=\ph$ where both 
   $x_{2\pi}$ and 
   $x_{g}{}\,(<x_{2\pi})$ become infinite. The lower limit is given by 
   $x_{g}=x_{0}$.}
  \label{fg:d1}
\efg

The next closed path will have the shape of an italic \textit{V} tilted to 
the left as in the inset of figure~\ref{fg:d3}. This graph has to be obtained 
numerically as described in the introduction to this section. 

\bfg
	\begin{center} 
	  {\scalebox{1}{\includegraphics
    {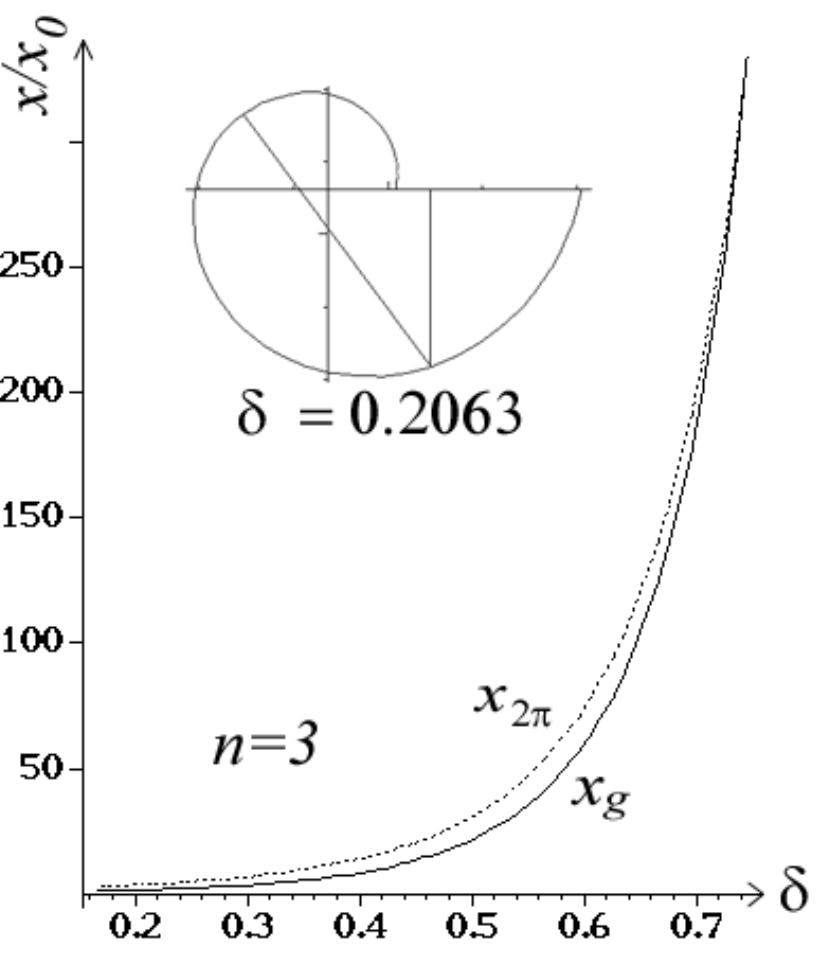}}}
	\end{center} 
  \caption{$x_{2\pi}$ and $x_{g}$ for a V-shaped degenerate closed 
   path with $n=3$ reflections on the \eas boundary. The inset shows such a 
   closed path for 
   $\delta = 0.2063$. The abscissa extends from $\delta_i$ to $\delta_f$ 
   as given in Table~\ref{tb:table1}.}
  \label{fg:d3}
\efg

The graph in figure~\ref{fg:d3} shows two curves. The upper one represents 
$x_{2\pi}/x_0$ and the lower one is $x_{g}/x_0$. It starts at 
$\delta_i\approx 0.1663186982 \,\,(x_{g} = x_0)$ and ends at 
$\delta_f \approx 0.7453672498$, i.e., the value of 
$\delta$ for which $x_{g} = 
x_{2\pi} \approx 330.053788$. For 
$\delta \notin [\delta_i,\delta_f]$ 
no such closed path will exist. As we can see in figure~\ref{fg:d3} the two 
curves for $x_g$ and $x_{2\pi}$ are rather close to each other. 
Figures~\ref{fg:d5},~\ref{fg:d7} and~\ref{fg:d9} show that this changes as 
the number of reflections in one cycle increases. Yet the other principal 
qualitative features like the limits to $\delta$, i.e., $\delta_i$ and 
$\delta_f$ with $x_g = x_0$ and $x_g = x_{2\pi}$, respectively, remain valid 
even for the paths with larger number $n$ of reflections at the 
\eas boundary. Table~\ref{tb:table1} lists the values of 
$\delta_i$, $\delta_f$ and $x_g = x_{2\pi}$ for $n=3,5,7$, and 9. 

\bfg
	\begin{center} 
	  {\scalebox{1}{\includegraphics
    {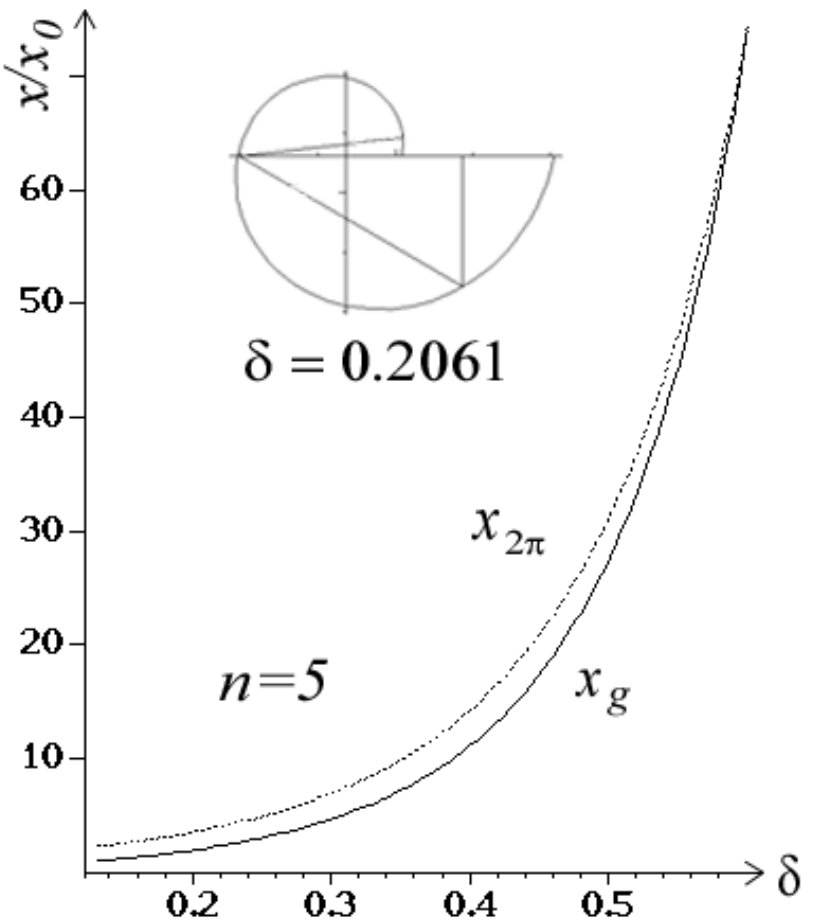}}}
	\end{center} 
  \caption{$x_{2\pi}$ and $x_{g}$ for a degenerate closed path with 
   $n=5$ reflections on the \eas boundary. The inset shows such a closed 
   path for 
   $\delta = 0.2061$. The abscissa extends from $\delta_i$ to $\delta_f$ 
   as given in Table~\ref{tb:table1}.}
  \label{fg:d5}
\efg

\bfg
	\begin{center} 
	  {\scalebox{1}{\includegraphics
    {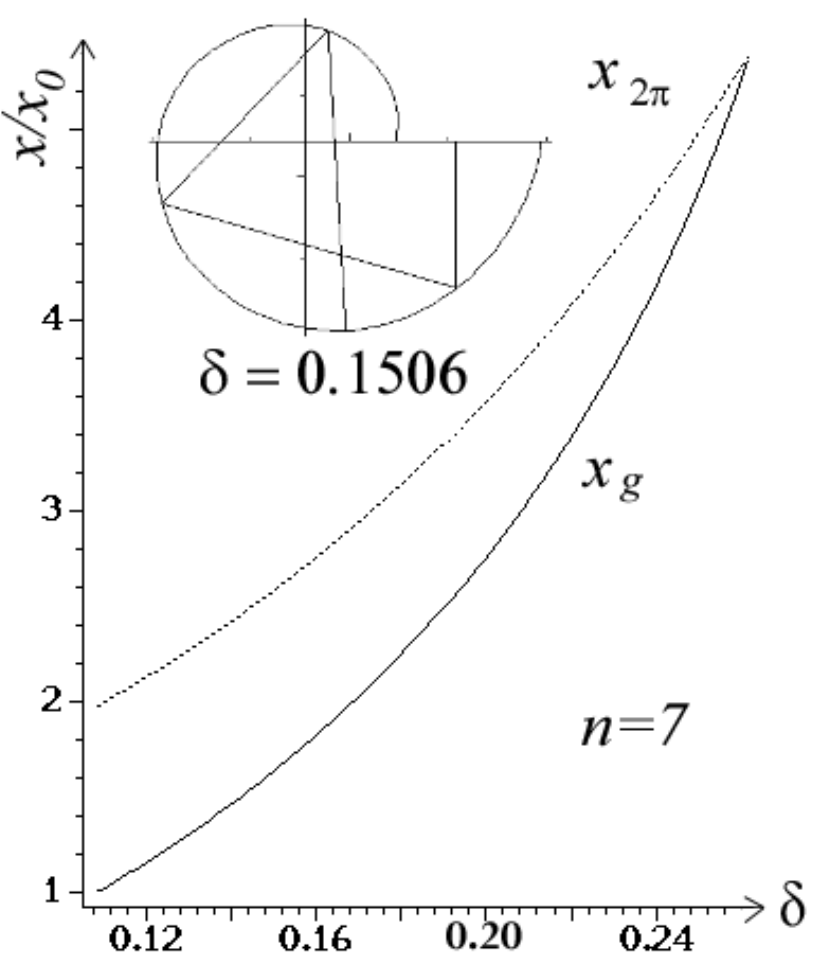}}}
	\end{center} 
  \caption{$x_{2\pi}$ and $x_{g}$ for a degenerate closed path with 
   $n=7$ reflections on the \eas boundary. The inset shows such a closed 
   path for 
   $\delta = 0.1506$. The abscissa extends from $\delta_i$ to $\delta_f$ 
   as given in Table~\ref{tb:table1}.}
  \label{fg:d7}
\efg

\bfg
	\begin{center} 
	  {\scalebox{1}{\includegraphics
    {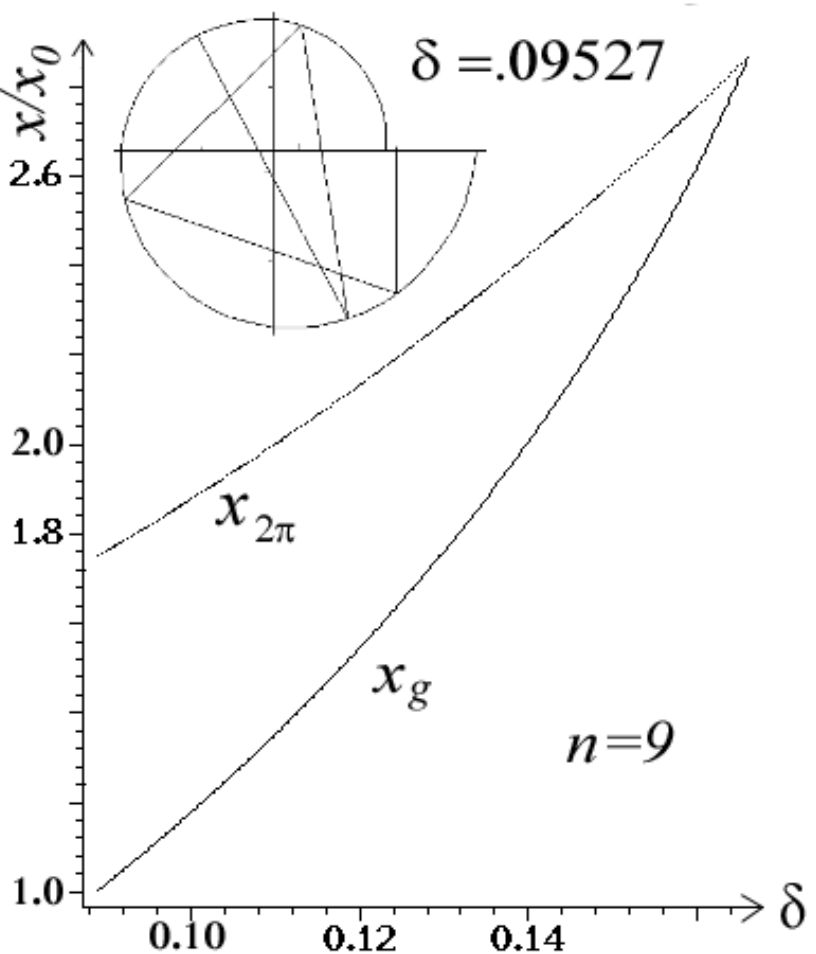}}}
	\end{center} 
  \caption{$x_{2\pi}$ and $x_{g}$ for a degenerate closed path with 
   $n=9$ reflections on the \eas boundary. The inset shows such a closed 
   path for 
   $\delta = 9.572 \times 10^{-2}$. The abscissa extends from $\delta_i$ 
   to $\delta_f$ as given in Table~\ref{tb:table1}.}
  \label{fg:d9}
\efg

\begin{table}[p]
\caption{$\delta_i$, $\delta_f$, and $x_g{}(\delta_f) = 
x_{2\pi}{}(\delta_f)$ for $n=3,5,7$, and 9. Values of $x_g$ in units of $x_0$.}
\label{tb:table1}
\begin{tabular}{ccccc}
\hline
\hline
& \multicolumn{4}{c}{$n$}
\\ 
       &\multicolumn{1}{c}{$3$} 
       &\multicolumn{1}{c}{$5$} 
       &\multicolumn{1}{c}{$7$} 
       &\multicolumn{1}{c}{$9$} 
\\ \hline 
$\delta_i$ & 0.1663186982
           & 0.1334789387
           & 0.1077909764
           & 0.08875113619
\\ 
$\delta_f$ & 0.7453672498
           & 0.6062445713
           & 0.2624642501
           & 0.1661601991
\\ 
$x_g = x_{2\pi}$  & 330.053788
                  & 77.97638
                  & 5.408937121
                  & 2.868332289 
\\  \hline \hline
\end{tabular}
\end{table}

Table~\ref{tb:table1} and its graphical rendering in figure~\ref{fg:table1} 
shows that $\delta_i$ and $\delta_f$ decrease with incerasing $n$, yet they 
still overlap each other mutually, except, e.g., for $n=3$ and $n=9$. 
This means that a kind of \textit{mode selection}\footnote{I use the term 
\textit{mode} loosely connected to the idea that a closed path may indicate 
the potential presence of a laser mode, given that the disk physically 
represents a laser medium. Compare also the discussion of orbits and 
resonator modes in~\cite{bowtie}.}

\bfg
	\begin{center} 
	  {\scalebox{1}{\includegraphics
    {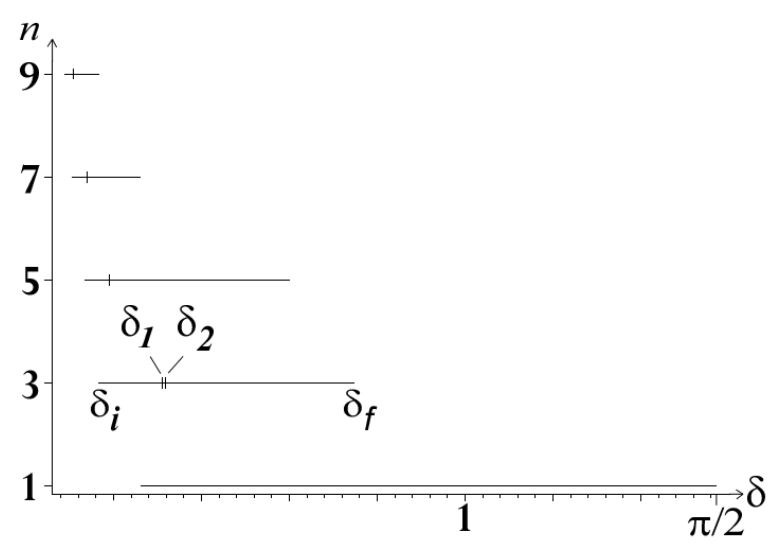}}}
	\end{center} 
  \caption{Deformation parameter $\delta$ intervals for the existence of 
   degenerate modes (long $[\delta_i,\delta_f]$ intervals) and nondegenerate 
   modes (short $[\delta_1,\delta_2]$ intervals), respectively. 
   For n=1 no nondegenerate 
   $[\delta_1,\delta_2]$ interval exists.}
  \label{fg:table1}
\efg

 is possible by choosing sufficiently 
high or sufficiently low values of $\delta$. E.g., $\delta = 0.7$ will 
select the modes with $n=1$, and $n=3$; or $\delta = 0.09$ will select 
the mode $n=9$, and most likely higher modes.

\subsection{Nondegenerate closed paths with $\varphi_1 = \varphi_n < \ph$}

\subsubsection{Nondegenerage closed paths with $n = 3$ reflections along 
the equiangular spiral boundary}

In the inset of figure~\ref{fg:n3} we see that the closed path with $n=3$ 
distinct nondengenerate points of reflection along the \eas boundary may 
be thought of as a splitting up of the degenerate path in figure~\ref{fg:d3}. 
It has the shape of an \textit{asymmetric bowtie}. Figure~\ref{fg:n3} 
shows three curves: $x_{2\pi}(\delta)$ (top), $x_g(\delta)$ (middle) and 
$\varphi_1(\delta)$ (bottom). 

\bfg
	\begin{center} 
	  {\scalebox{1}{\includegraphics
    {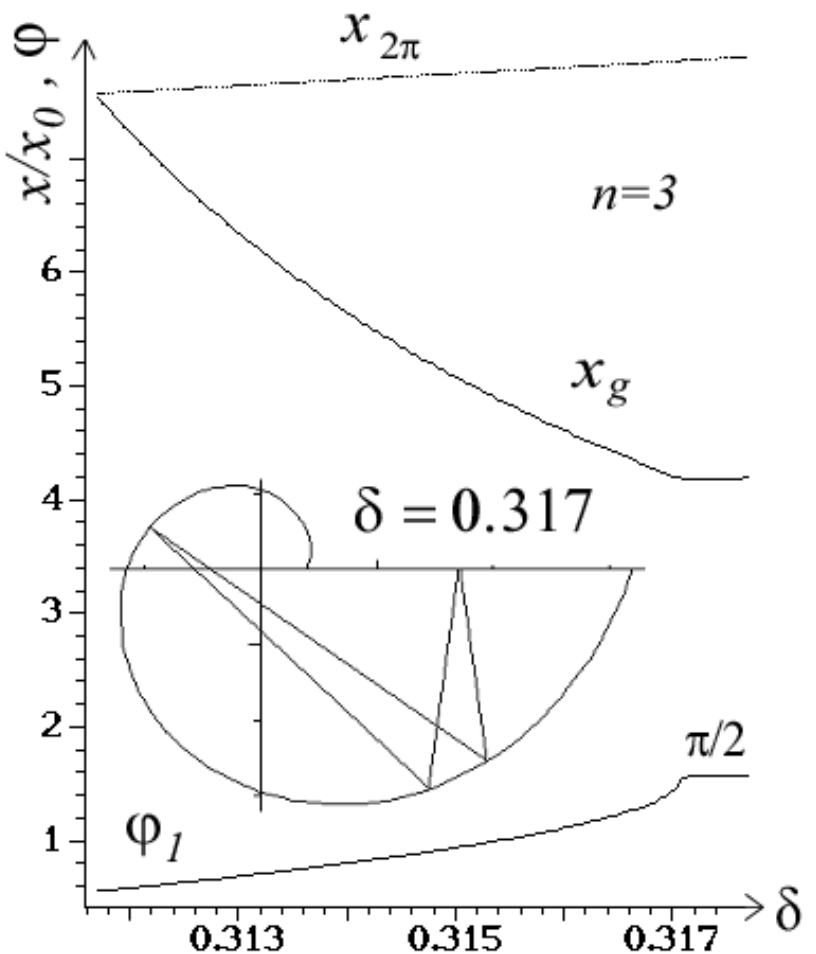}}}
	\end{center} 
  \caption{$x_{2\pi}$, the point of reflection on the gap $x_{g}$ and 
   the angle $\varphi_1$ (compare figure~\ref{fg:ex9}) for the (supposedly) 
   simplest nondegenerate closed path with $n=3$ reflections on the \eas 
   boundary. It has the shape of an \textit{asymmetric bowtie} as shown 
   in the inset for $\delta=0.317$. For $\delta > \delta_2$, as given in 
   Table~\ref{tb:table2}, the closed path becomes degenerate 
   (comp. figure~\ref{fg:d3}) and $\varphi_1 \equiv \ph$.}
  \label{fg:n3}
\efg

$x_g$ becomes equal to $x_{2\pi}$, if 
$\delta = \delta_1 \approx 0.3116739102$. $\varphi_1(\delta)$ is a 
monotonous increasing function which becomes equal to 
\phtext~at 
$\delta = \delta_2 \approx 0.31710313$. For $\delta > \delta_2$ only the 
degenerate closed path with $n=3$ continues to exist. Comparing the intervals 
$[\delta_1,\delta_2]$ and $[\delta_i,\delta_f]$ of the last section we see 
that $[\delta_1,\delta_2] \subset [\delta_i,\delta_f]$. Hence both, 
degenerate and nondegenerate closed paths with $n=3$ reflections on the 
\eas boundary coexist for all values $\delta \in [\delta_1,\delta_2]$.

Figure~\ref{fg:stable} shows the four angles of incidence 
$\gamma_k{}(\delta) = \,\,\abs{\ta_k{}(\delta) + \delta} \,\, (k=1,2,3)$, and 
$\gamma_4{}(\delta) = \varphi_4(\delta) [=\varphi_1(\delta)] $ 
relative to directions perpendicular to the boundary at 
\vind{x}{k} $(k=1,2,3)$, and \vind{x}{g}, respectively, for the interval 
$\delta \in [\delta_1,\delta_2]$. 
The horizontal line depicts the critical angle of total reflection $\gamma_c$,
calculated from 
\be
  \sin \gamma_c = 1/n_r
  \label{eq:totref}
\ee
for $n_r=3.3$. This is the refractive index of the cascaded InGaAs/InAlAs
 system used in the well-known symmetric bow-tie micro-disk laser 
experiments~\cite{bowtie}. Figure~\ref{fg:stable} shows that for a 
small intervall 
around $\delta = 0.3152 \in [\delta_1,\delta_2]$ all four angles of incidence 
$\gamma_k, \,\,(k=1,\ldots, 4) $ are greater than the critical angle of 
total reflection $\gamma_c$. 

\bfg
	\begin{center} 
	  {\scalebox{1}{\includegraphics
    {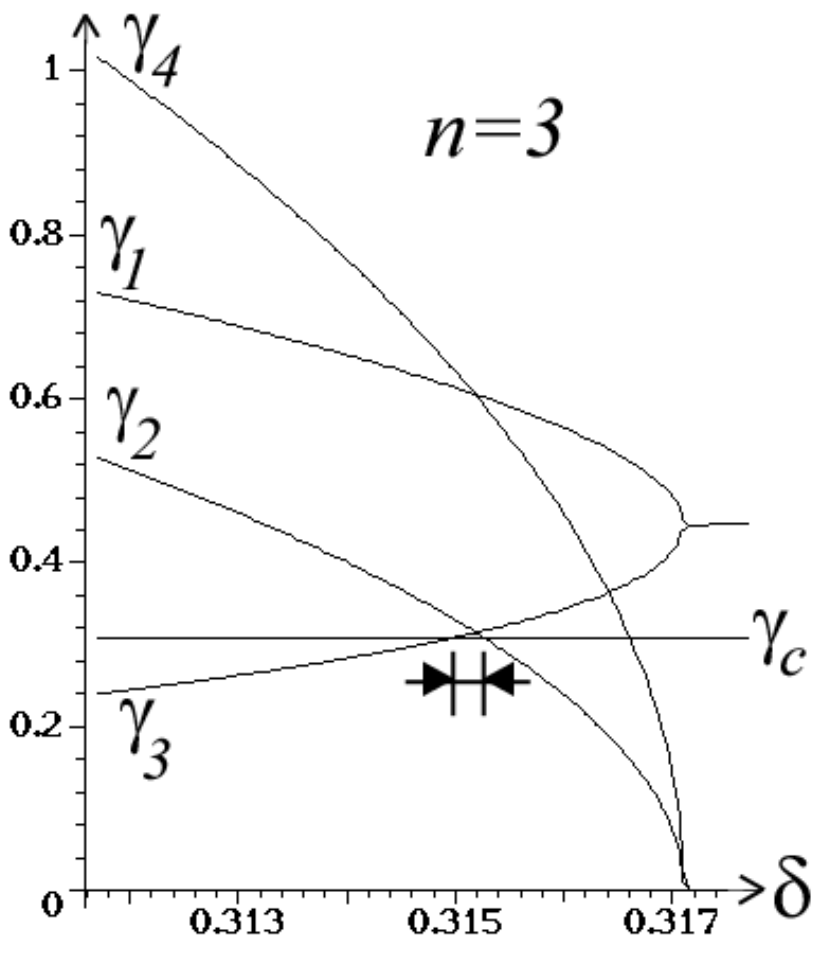}}}
	\end{center} 
   \caption{Angles of incidence $\gamma_k, \,\, (k=1,\ldots,4)$ at 
   \vind{x}{1}, \vind{x}{2}, \vind{x}{3}, and \vind{x}{g}, respectively, 
   for the asymmetric 
   bow-tie shaped closed paths with $n=3$ incidences on the \eas boundary. 
   Abscissa: $\delta_1 < \delta < \delta_2$ (comp. Table~\ref{tb:table2}).
   Critical angle for total reflection $\sin\gamma_c = 1/n_r$ 
   with $n_r = 3.3$. Especially marked is the narrow $\delta$-interval 
   for which all $\gamma_k > \gamma_c$.}
   \label{fg:stable}
\efg

This means that light following the $n=3$ nondegenerate closed path
 trajectories, i.e., the \textit{asymmetric bow-tie} trajectories as depicted 
in the inset of figure~\ref{fg:n3}, will stay ``forever'' trapped within 
the \eas micro-disk resonater (neglecting evanescent leakage of tunneling 
of photons~\cite{bowtie}). This will be the case, provided that the deformation
parameter $\delta$ is contained in the small intervall around $0.3152$ 
mentioned above, such that $\gamma_k > \gamma_c, \,\, (k=1,\ldots,4)$. 
I therefore expect the \textit{asymmetric bow-tie} 
mode introduced here to be able to lase.\footnote{The same condition was 
necessary for the symmetric bow-tie mode to be able to lase~\cite{bowtie}.}

By virtue of its asymmetry, I expect a further increase in the directionality 
of the power output, which is to be highest in \textit{one} of the four 
directions pertaining to \vind{x}{1}, \vind{x}{2}, \vind{x}{3} and 
\vind{x}{g}; compared to the symmetric bow-tie mode with equal power 
output in all four directions~\cite{bowtie}.

\subsubsection{Nondegenerage closed paths with $n \geq 5$ reflections 
along the equiangular spiral boundary}

Comparing Figs.~\ref{fg:n3},~\ref{fg:n5},~\ref{fg:n7}, and~\ref{fg:n9} 
shows that the basic features, which have just been explained for the case of 
$n=3$ nondegenerate reflections continue to persist also for higher values of 
$n$. Only the transition in the $\varphi_1$ curve at $\delta = \delta_2$ 
seems to become more and more abrupt.

\bfg
	\begin{center} 
	  {\scalebox{1}{\includegraphics
    {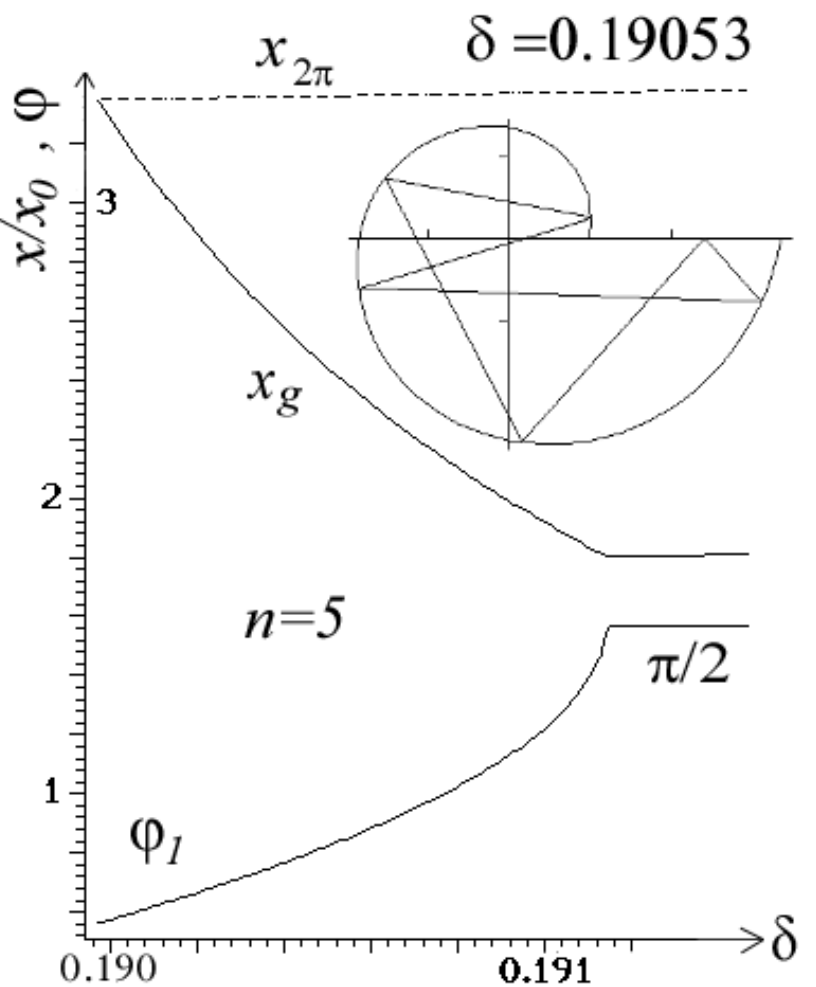}}}
	\end{center} 
  \caption{$x_{2\pi}$, the point of reflection on the gap $x_{g}$ and 
   the angle $\varphi_1$ (compare figure~\ref{fg:ex9}) for the nondegenerate 
   closed path with $n=5$ reflections on the \eas boundary. An example 
   of such a closed path is shown in the inset for $\delta=0.19053$. 
   For $\delta > \delta_2$ (comp. Table~\ref{tb:table2}) 
   the path becomes degenerate as in figure~\ref{fg:d5} and 
   $\varphi_1 \equiv \ph$.}
  \label{fg:n5}
\efg

\bfg
	\begin{center} 
	  {\scalebox{1}{\includegraphics
    {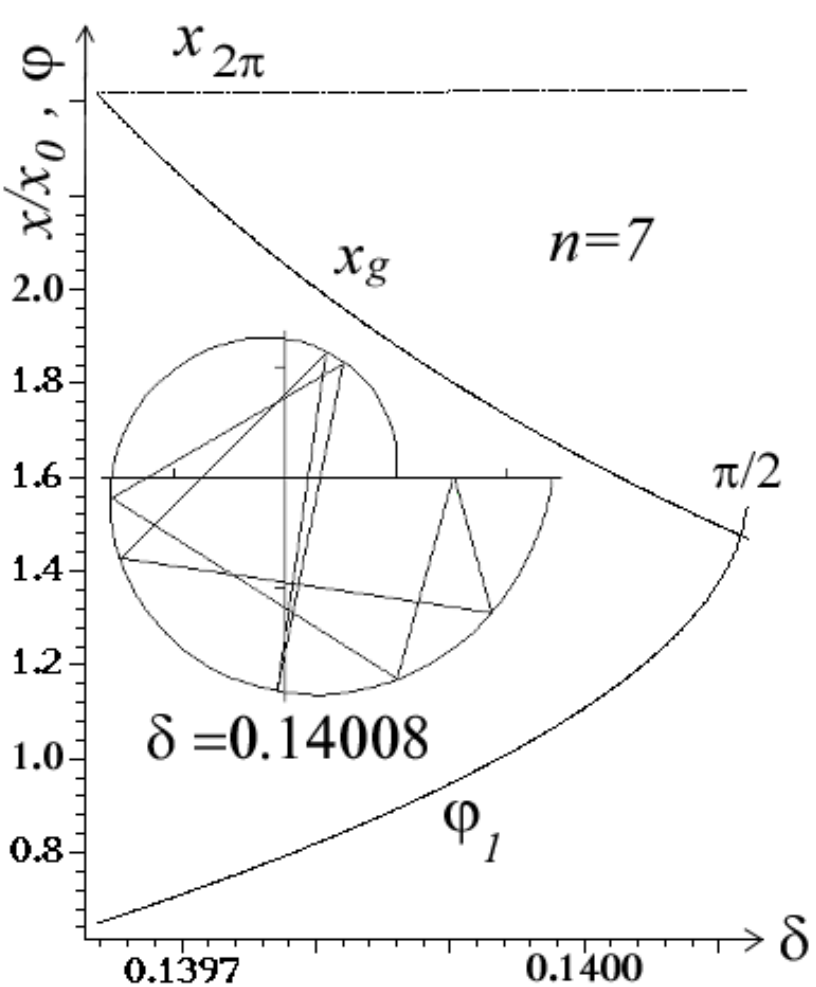}}}
	\end{center} 
  \caption{$x_{2\pi}$, the point of reflection on the gap $x_{g}$ and 
   the angle $\varphi_1$ (compare figure~\ref{fg:ex9}) for the nondegenerate 
   closed path with $n=7$ reflections on the \eas boundary. An example of 
   such a closed path is shown in the inset for $\delta=0.14008$. 
   For $\delta > \delta_2$ (comp. Table~\ref{tb:table2}) the path 
   becomes degenerate as in figure~\ref{fg:d7} and 
   $\varphi_1 \equiv \ph$.}
  \label{fg:n7}
\efg

\bfg
	\begin{center} 
	  {\scalebox{1}{\includegraphics
    {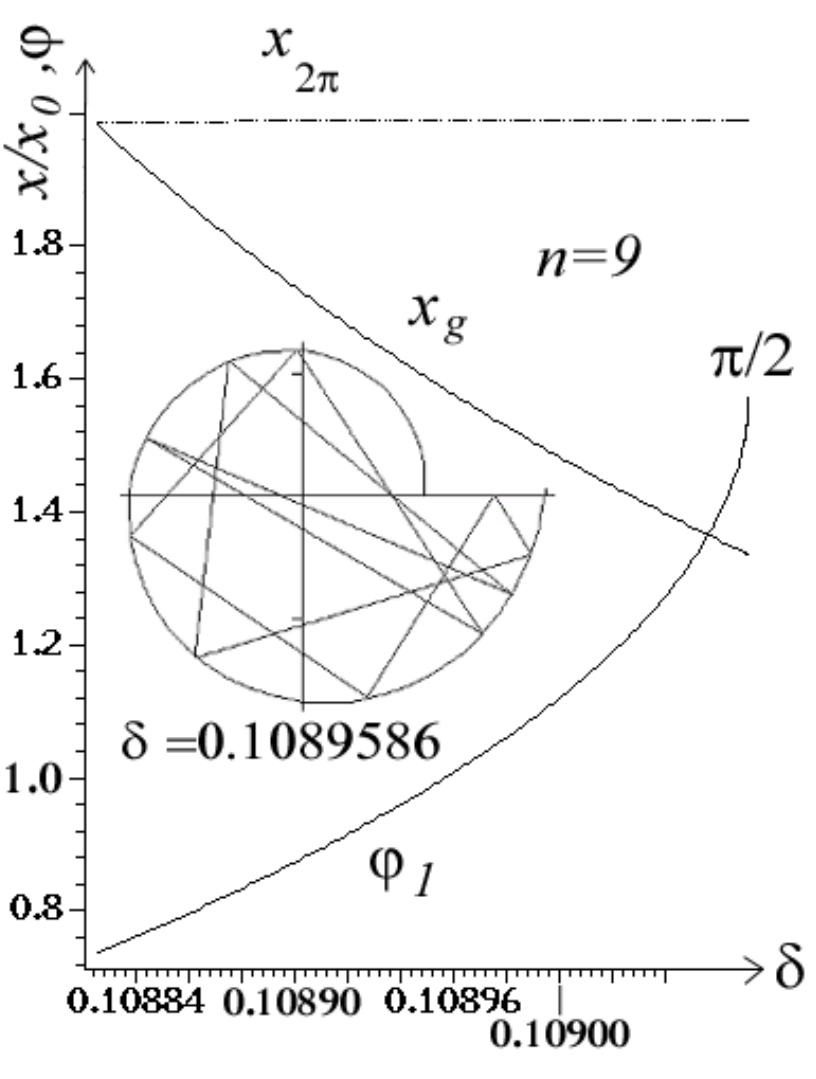}}}
	\end{center} 
  \caption{$x_{2\pi}$, the point of reflection on the gap $x_{g}$ and 
   the angle $\varphi_1$ (compare figure~\ref{fg:ex9}) for the nondegenerate 
   closed path with $n=9$ reflections on the \eas boundary. 
   An example of such a closed path is shown in the inset for 
   $\delta=0.1089586$. For $\delta > \delta_2$ 
   (comp. Table~\ref{tb:table2}) the path becomes degenerate as in 
   figure~\ref{fg:d9} and $\varphi_1 \equiv \ph$.}
  \label{fg:n9}
\efg

Table~\ref{tb:table2} lists the values of $\delta_1$, $x_g(\delta_1)=
x_{2\pi}(\delta_1)$, $\varphi_1(\delta_1)$, $\delta_2$, $x_g(\delta_2)$, 
and 
$(\delta_f-\delta_i)/(\delta_2-\delta_1)$ 
for $n=3,5,7$, and 9. For the case of odd $n$ the condititon $x_g(\delta_1)=
x_{2\pi}(\delta_1)$ is equivalent to:
\bdm
  \delta_1 = -\frac{3}{2}\pi + \sum_{k=1}^{(n-1)/2} (\Delta_{2k+1,2k}-\pi)
\edm

\begin{table}[p]
\caption{$\delta_1$, $x_g(\delta_1)=x_{2\pi}(\delta_1)$, 
$\varphi_1(\delta_1)$, $\delta_2$, $x_g(\delta_2)$, and 
$(\delta_f-\delta_i)/(\delta_2-\delta_1)$ for 
$n=3,5,7$, and 9. Values of $x_g$ in units of $ x_0 $.}
\label{tb:table2}
\begin{tabular}{ccccc}
\hline \hline
& \multicolumn{4}{c}{$n$}    
\\
       &\multicolumn{1}{c}{3} 
       &\multicolumn{1}{c}{5} 
       &\multicolumn{1}{c}{7} 
       &\multicolumn{1}{c}{9} 
\\ \hline 
$\delta_1$ & 0.3116739102
           & 0.1899678060
           & 0.1396366678
           & 0.1088242806
\\ 
$x_g = x_{2\pi}$ 
           & 7.5706580
           & 3.3473581
           & 2.41840995
           & 1.98670040
\\ 
$\varphi_1(\delta_1)$  
                  & 0.551868221
                  & 0.561810573
                  & 0.648663822
                  & 0.734333562
\\ 
$\delta_2$        & 0.31710313
                  & 0.1911406275
                  & 0.14012182865
                  & 0.10907108189
\\ 
$x_g(\delta_2)$ 
           & 4.169707
           & 1.8034054
           & 1.469010
           & 1.3357671
\\ 
$(\delta_f-\delta_i)/(\delta_2-\delta_1)$ 
           & 107
           & 403
           & 319
           & 314
\\  \hline \hline
\end{tabular}
\end{table}

The quotient of the length of the degenerate and the nondegenerate $\delta$ 
intervals in Table~\ref{tb:table2} shows that the nondegenerate interval 
lengths are all less than 1\% of the degenerate ones. This partly explaines 
why the nondegenerate data are more difficult to be calculated in the first 
place. Table~\ref{tb:table2} or figure~\ref{fg:table1} clearly show that for a 
given $\delta$ of the nondegenerate modes with $n=3,5,7$, and 9, if at all, 
only one of these nondegenerate modes will exist. The reason is that the 
$[\delta_1,\delta_2]$ intervals don`t overlap. This makes a unique 
selection amongst the nondegenerate modes possible by simply choosing an 
appropriate $\delta$ from one of the 
$[\delta_1,\delta_2]$-intervals, if we neglect the degenerate modes for the 
moment being. 

Table~\ref{tb:table2} suggests further that the center points and the lengths 
of the
 $[\delta_1,\delta_2]$ intervals decrease continually with increasing $n$, 
whereas the lengths proportions of the nondegenerate and degenerate $\delta$ 
intervals pertaining to the same $n$ may continue to be of an order of less 
than 1\%. 
It is therefore likely that the above suggested unique selection of 
nondegenerate modes may continue to work for higher $n$ as well. 

The fact that all $[\delta_1,\delta_2]$ intervals for nondegenerate modes with
$n \geq 5$ have $\delta_2 < \gamma_c(n_r=3.3)$ 
shows that for refractive indices 
$n_r \leq 3.3$ no further \textit{stable} nondegenerate closed path light 
trajectories are to be expected. E.g., for $n=5$ the refractive index would 
in principle have to be as high\footnote{Calculated according to 
Eq. (\ref{eq:totref}) with 
$\delta_2(n=5)$ from Table~\ref{tb:table2}.}
as $n_r=5.26$, in order to create a situation with the possibility of all 
angles of incidence to be greater than $\gamma_c$.

\subsection{Existence of nondegenerate modes with even $n$}

So far I did not find a numerical solution for a nondegenerate mode with an 
even number $n$ of reflections along the \eas boundary, when applying the 
same techniques as just explained for the case of odd $n$. In the case of 
odd $n$, a nondegenerate mode may be thought to naturally arise from the 
splitting up of the corresponding degenerate mode as the branching off of 
the nondegenerate mode in Figs.~\ref{fg:n3}, and \ref{fg:n5} to~\ref{fg:n9} 
suggests. Such a branching off into a nondegenerate mode with even $n$ 
seems unlikely, since no corresponding degenerate mode with an even number 
of reflections $n$ on the \eas boundary exists. This consideration may serve 
as a hint, but seems by itself certainly insufficent to explain the possible 
nonexistence of nondegenerate modes with even $n$.

\section{Conclusion}

A new type of deformation, the \eas deformation of circular disks has been 
introduced. This deformation simultaneously breaks the symmetry of rotation 
and contains no remaining symmetries of reflection as in the case of the 
flattened quadrupoles or ovals~\cite{bowtie, Phot-Bill}. A combination of 
\eas and oval deformations may therefore be a promising direction for future 
investigations. 

The real two-dimensional geometric calculus description of equiangular 
spirals was explained briefly together with a short review of the geometric 
properties of light ray propagation inside an equiangular spiral.

Then the constituting equations for closed light paths inside an \eas were 
derived and their explicite forms fit for numerical solution were given. 
The numerical solution of these constituting equations showed that there 
are basically two generic types of closed paths: degenerate paths with 
vertical incidence at the gap and nondegenerate paths with nonvertical 
incidence at the gap. The nondegenerate paths were seen to exist only for 
very specific intervals of the deformation parameter $\delta$ less than a 1\% 
fraction of the permissible $\delta$-intervals of the degenerate closed 
paths. All interval lenghts, centerpoints and boundaries monotonically 
decreased with increasing $n$ as shown in figure~\ref{fg:table1}.

It is to be expected that for reflectivities $R$ less than unity, e.g., 
for dielectrica with $R=(n_r-1)^2/(n_r+1)^2$ (Fresnel result for vertical 
incidence~\cite{Phot-Bill, Jackson}) degenerate modes with vertical incidence 
at the gap may not be able to lase. Hence a choice of deformation parameter 
$\delta$ in a $[\delta_1,\delta_2]$ interval specific to some value of $n$ 
will enable the exclusive selection of the presence of a nondegenerate mode 
for this particular \eas disk. 

It was further shown that equiangular spirals with deformation parameters 
contained in a narrow interval around $\delta=0.3152$, exhibit \textit{stable 
asymmetric bow-tie} shaped light trajectories, assuming a high refractive 
index material with $n_r=3.3$ as in~\cite{bowtie}. Compared to the symmetric 
bow-tie micro-disk laser~\cite{bowtie} an increase in the directionality of 
the output power in only \textit{one} preferred direction is to be expected. 

In future investigations the Poincar\'{e} sections recording the angles of 
incidence $\gamma=\,\,\abs{\ta+\delta}$ versus the polar angle 
$\Phi$ should be 
examined for stable orbits above the total reflectivity escape condition 
(\ref{eq:totref}). This would give sufficiently strong reasons for 
endeavouring full numeric solutions of the Helmholtz wave equation and 
for conducting corresponding experiments.

\section*{Acknowledgements}

I express my thanks to God by quoting from Genesis: ``In the beginning God 
created the heavens and the earth...And God said, 
`Let there be light,' and there was light.~\cite{NIV:light}''
I thank Dr. H. Ishi from Yokohama City University, Japan, for stimulating 
discussions, and I thank Prof. H. Dehnen from Konstanz University, Germany, 
for his helpful comments and suggestions. I finally express my gratitude 
towards Fukui University for providing a stuitable environment to conduct 
this research.


\clearpage
\pagestyle{empty}

\end{document}